\documentclass{new_tlp} 
\usepackage{amsmath}
\usepackage{amssymb}
\usepackage{mathptmx}
\usepackage{hyperref}
\usepackage[titletoc]{appendix}
\usepackage{yfonts}
\usepackage{listings}
\newtheorem{example}{Example}[section]
\newtheorem{lemma}{Lemma}[section]
\newtheorem{theorem}{Theorem}[section]
\newtheorem{corollary}{Corollary}[section]
\newcommand\round[1]{\left[#1\right]} 
\newcommand{\myHat}{{\hat{\hspace{0.5em}}}}
\title[Coinduction Plain and Simple]{Coinduction Plain and Simple}

  \author[Fran\c{c}ois Bry]
         {FRAN\c{C}OIS BRY\\
         Institute for Informatics, Ludwig-Maximilian University of Munich, Germany\\
         \email{bry@lmu.de}
         }

\jdate{July 2020}

\pagerange{\pageref{firstpage}--\pageref{lastpage}}
\doi{S1471068401001193} 
\newtheorem{definition}{Definition}[section]
\newtheorem{proposition}{Proposition}[section]
 
\begin{document}

\label{firstpage}

\maketitle  

\begin{abstract}
Coinduction refers to both a technique for the definition of infinite streams, so-called codata, and a technique for proving the equality of coinductively specified codata.
This article first reviews coinduction in declarative programming. Second, it reviews and slightly extends the formalism commonly used for specifying codata. Third, it generalizes the coinduction proof principle, which has been originally specified for the equality predicate only, to other predicates. This generalization makes the coinduction proof principle more intuitive and stresses its closeness with structural induction. The article finally suggests in its conclusion extensions of functional and logic programming with limited and decidable forms of the generalized coinduction proof principle. 
\end{abstract}


\tableofcontents

\section{Introduction}\label{sec:Introduction}

A \emph{data stream}, short \emph{stream}, is a possibly never ending sequence of data. A stream is \emph{observed} when its data result from observations or measurements of natural or artificial systems. An observed data stream might for example convey temperatures, the reproduction rates of a disease, or energy consumption or traffic volumes. A stream is \emph{constructed} when its data are synthetic. Constructed streams are useful as encoded models of observed streams.
Using streams, an early detection of critical situations can be specified as the continuous comparison of the data of an observed stream with that of a constructed stream modelling some behavior.  

Constructed streams are finitely defined from non-stream data and operations on such data. Thus, two disjoint data types are considered in the following: Non-stream data the type of which is called \emph{data} and stream data the type of which is called \emph{codata}.

Constructed streams, or codata, are finitely defined for, otherwise, they could not be used in programming. Codata express repetitions by recursion. As a consequence, recursive functions like the Fibonacci function can be expressed as codata. The specification of recursive functions as codata is interesting in its own right, independently of stream-related applications, because recursive codata yield iterative computation processes like tail recursive functions do but without the recourse to accumulators. 

Streams are related to processes. An observation process operating on an observed stream might be unpredictable, or only partly predictable, while a process operating on a constructed stream, even though it might never end, can be given a finite definition derived from that of the constructed stream it refers to. Coinduction in declarative programming has been used for the specification of both constructed streams and processes operating on constructed streams. 

This article first reviews coinduction in declarative programming. Second, it reviews and slightly extends the formalism commonly used for specifying codata. Third, it generalizes the coinduction proof principle to other predicates than equality. The article finally suggests in its conclusion extensions of functional and logic programming with limited and decidable forms of the generalized coinduction proof principle. 

\section{Coinduction in Declarative Programming}
\label{set:Coinduction-in-Programming-so-far}

Coinduction was introduced in 1971 by Robin Milner in an investigation of the correctness of terminating or non-terminating imperative programs \cite{Milner-Simulation-1971}. More precisely, the article \cite{Milner-Simulation-1971} introduced simulation relations and a precursor of the coinductive proof principle consisting in exhibiting finitely many pairs of related states and verifying that a simulation relation holds of each pair. David Park defined in 1981 bisimilarity as a greatest fixpoint, derived from fixpoint theory the bisimulation proof method \cite{Park-Concurrency-on-automata-and-infinite-sequences-1981} which is now also called coinductive proof principle. Bisimulation and the coinductive proof principle soon became cornerstones of process algebra \cite{Hoare-CSP-1978,Fokkink-Introduction-to-process-algebra-1999} and of the theory of communicating concurrent processes \cite{Milner-CCS-1980}. Davide Sangiorgi gives in \cite{Sangiorgi-Origins-of-bisimulation-and-coinduction-2009} an extensive account of the origins of coinduction, bisimulation and of related precursor concepts and methods in philosophical and mathematical logic, mathematics and computer science which, however, does not address coinduction in declarative programming and does not mention inductionless induction \cite{Comon-Inductionless-Induction-2001}. 

Coinductive definitions, or codata, have been introduced in 1982 by Alain Colmerauer in logic programming under the name of \emph{rational trees} \cite{colmerauer-prolog-and-infinite-trees-1982}, ``tree'' meaning finite or infinite term. A rational tree is a finite or infinite tree the set of subtrees of which is finite. The name ``rational tree'' which,  puzzlingly, is not explained in \cite{colmerauer-prolog-and-infinite-trees-1982,colmerauer-equations-inequations-on-infinite-trees-1984}, was chosen in reference to the fact that a real number is a rational number if and only if its decimal expansion is finite or repeating \cite{colmerauer-private-communication-1988}. The name ``rational tree'' appropriately stresses that, by Georg Cantor's diagonal argument \cite{Cantor-Diagonalization-1891}, the set of (rational and non-rational) trees over a finite or denumerable alphabet is not denumerable\footnote{This fact challenges model theories for logic programming with observed streams.} while the set of rational trees of course is denumerable. 

Following Courcelle \cite{courcelle-fundamental-properties-of-infinite-trees-1983}, rational trees are also named \emph{regular trees}. In the 1980s rational trees were easily expressed in Prolog because, for efficiency reasons at a time processors were slow, Prolog did not perform the occurs-check during unification: An infinite list of $1$s could for example be finitely expressed as \verb#X = [1|X]# or, redundantly, as \verb#X = [1,1,1|X]# and an infinite list of alternating $0$s and $1$s as \verb#X = [0,1|X]#. Colmerauer systematized this observation and designed a variant of Prolog, Prolog II \cite{colmerauer-equations-inequations-on-infinite-trees-1984,van-Emden-A-logical-reconstruction-of-PrologII-1983}, which supported rational trees. Michael Maher gave in 1988 in \cite{Maher-complete-axiomatization-rational-trees-1988} an axiomatization of the algebras of rational trees. Unification algorithms for rational trees were proposed in \cite{mukai-unification-algorithm-for-infinite-trees-1983,Jaffar-Efficient-Unification-Over-Infinite-Terms-1984,Martelli-Rossi-Unification-with-Infinite-Trees-1984}. A meta-interpreter for a logic programming language supporting rational (or regular) trees (or terms) is described in \cite{ancona-regular-corecursion-in-prolog-2013}.\footnote{The article does not clarify in which cases unification is performed with, respectively without, occurs-checks.} 

Coinductive definitions in logic programming are possible not only with rational trees defined using unification without occurs-check, but also through non-terminating SLD derivations based on unification with occurs-check. This approach has been first investigated in 1983 by Nait Abdallah in \cite{Abdallah-Metric-Interpretations-and-Greatest-Fixpoint-Semantics-of-LP-1983}, further investigated in \cite{Abdallah-Interpretation-of-Infinite-Computations-in-LP-1984,van-Emden-Top-down-semantics-1985,Palamidessi-Formal-Semantics-1985}\footnote{The ``completion of a program'' mentioned in \cite{van-Emden-Top-down-semantics-1985} is Clark's completion semantics of negation in logic programming \cite{Clark-Negation-as-Failure-1978}. Thus, it is not related to the metric-based completion of the Herbrand universe introduced in \cite{Abdallah-Metric-Interpretations-and-Greatest-Fixpoint-Semantics-of-LP-1983}.} and popularized in 1987 by John Lloyd with the chapter on perpetual processes of the second edition of his Foundations of Logic Programming \cite{Lloyd-Foundations-of-Logic-Programming-1987}. Each stage of a non-terminating SLD derivation like that resulting from the evaluation of \verb#?- p(Y).# against the clause \verb#p(X) :- p(f(X)).# is seen as a finite approximation of an atom containing an infinite term \verb#f(f(f(...)))#. Nait Abdallah formalized such infinite terms as follows.

The Herbrand universe $H$ is first equipped with a metric: 
\begin{quote}
\begin{tabular}{l l l l}
   $d(t_1, t_2)$ 	& $=$ 	& $0$ 								& if 	$t_1 = t_2$ \\
   $d(t_1, t_2)$	& $=$ 	& $2^{-\text{inf}\{n | a_n(t_1){\neq}a_n(t_2)\}}$ 	& otherwise
\end{tabular}
\end{quote}
where $a_n(t)$ is the ``cut at depth n'' of a term $t$. Thus, $\text{inf}\{n | a_n(t_1) \neq a_n(t_2)\}$ is the smallest depth at which $t_1$ and $t_2$ differ.
Using this metric, Cauchy sequences of elements of $H$ are defined,\footnote{A Cauchy sequence is such that the distance between two successive elements tends to zero when the rank tends to infinity.} the standard equivalence relation on Cauchy sequences is considered,\footnote{Two Cauchy sequences are equivalent if the difference between their elements of same ranks tends to zero when the rank tends to infinity.} and the \emph{complete Herbrand universe} $\overline{H}$ is defined as the set of equivalence classes of Cauchy sequences of elements of $H$. Thus, $H \subset \overline{H}$.\footnote{Like $\mathbb{Q} \subset \overline{\mathbb{Q}} = \mathbb{R}$.} 
It is then observed that, by a result of \cite{Mycielski-Taylor1976_Article_ACompactificationOfTheAlgebra}, $\overline{H}$ is compact \footnote{The compactness of $\overline{H}$ means that every Cauchy sequence of elements of $\overline{H}$ has a limit in $\overline{H}$.} if its terms are defined from finitely many symbols. This is the case of the ground (or closed) terms of $\overline{H}$  because programs are finite. Thus, an infinite sequence of elements of $H$ like for example $f^n(a)$ has a limit in $\overline{H}$ which is denoted $f(f(\ldots))$ or $f(\omega)$.

The \emph{complete Herbrand base} $\overline{B}$ is defined as the set of ground atoms built from terms in the complete Herbrand universe $\overline{H}$, an immediate consequence operator $T'$ is defined in reference to the complete Herbrand base $\overline{B}$ in the same manner as the standard immediate consequence operator $T$ is defined in reference to the Herbrand base $B$ \cite{van-Emden-Kowalski-Semantics-of-Predicate-Logic-1976}. The greatest fixpoint semantics of a definite (that is, negation-free) logic program is defined as the greatest fixpoint of $T'$ for that program.
The greatest fixpoint semantics of logic programs is extended to programs with negation in \cite{Jaffar-Stuckey-Semantics-of-Infinite-Tree-Logic-Programming-1986,Hein-Completions-of-perpetual-logic-programs-1992}. 
Refinements of the greatest fixpoint semantics of logic programs are proposed in \cite{Golson-Declarative-Semantics-for-Infinite-Objects-in-LP-1988,Levi-Palamidessi-Contributions-To-The-Semantics-Of-Logic-Perpetual-Processes-1988,Jaume-On-Greatest-Fixpoint-Semantics-of-LogicProgramming-2002,Ancona-extending-co-logic-programming-with-cofacts-2017,Komendantskaya-Productive-Corecursion-in-Logic-Programming-2017,Li-Models-2018}. 


Some approaches to coinduction in logic programming make use of ancestor subsumption \cite{Socher-Ambrosius-Ancestor-subsumption-1992}, a technique for avoiding redundant derivations in resolution theorem proving: If a resolvent $R$ is subsumed (without occurs-check) by one of its ancestors $A$, then further derivations from $R$ can be avoided without compromising completeness. Indeed, if $R$ is subsumed (without occurs-check) by $A$, then every derivation from $R$ is also possible from $A$ and therefore redundant. 

In the 1980s, as coinductive definitions began to be considered in logic programming, it was folklore knowledge in the community that ancestor subsumption  (without occurs-check) can be used for preventing redundant non-terminating SLD derivations (with occurs-check) like in the following examples:\footnote{Ancestor subsumption was at the time a commonly considered redundancy elimination technique in resolution theorem proving.} 

\begin{example}
\label{ex:P1}
\begin{verbatim}
 
   p(0).
   allp([H|T]) :- p(H), allp(T). 
   
   ?- X = [0|X], allp(X). 
\end{verbatim}
\end{example}

\begin{example}
\label{ex:P2}
\begin{verbatim} 

   p :- p.
   q(X) :- q(X).
   r(f(X)) :- r(f(X)). 
   s(X) :- s(f(X)). 
   
   ?- p.
   ?- q(Y).
   ?- r(Y). 
   ?- s(Y). 
\end{verbatim}
\end{example}

\begin{example}
\label{ex:P3}
\begin{verbatim} 

   member(H, [H|_]).
   member(X, [_|T]) :- member(X, T). 
   
   ?- T = [0|T], member(1, T). 
\end{verbatim}
\end{example}

Avoiding redundant derivations by ancestor subsumption  (without occurs-check) or any other means does not mean resolving, though, as the examples given above illustrate. 

In Example \ref{ex:P1}, \verb# X = [0|X]# refers to unification without occurs-check and 
ancestor subsumption avoids that a Prolog system not performing the occurs-check repeatedly proves \verb#p(0)# and tries to prove \verb#allp(X)#. Without reasoning by induction, the proof of the goal \verb#X = [0|X],# \verb#allp(X)# cannot be completed. By induction, it can of course be proved. 

In the Example \ref{ex:P2}, ancestor subsumption  (without occurs-check) avoids that a Prolog system (performing the occurs-check or not) endlessly attempts to prove \verb#p#,  \verb#q(Y)#, \verb#r(f(Y))#, and \verb#s(f(...f(Y)...))#. Even by induction, none of these four goals can be proved. 

Example \ref{ex:P3} is similar to Example \ref{ex:P2}: With ancestor subsumption (without occurs-check), a Prolog system not performing the occurs-check does not endlessly attempt to prove \verb#member(1,# \verb#T)#.  Even by induction, the goal  \verb#member(1, T)# cannot be proved because \verb#1# does not occur in the cyclic list specified by \verb#T = [0|T]#.
 
Summing up, Examples \ref{ex:P1}, \ref{ex:P2} and \ref{ex:P3} illustrate that ancestor subsumption  (without occurs-check) \cite{Socher-Ambrosius-Ancestor-subsumption-1992} is a pruning rule (that is, a rule for discarding redundant parts of a proof), not an inference rule (that is, a rule for deriving logical consequences). 

The downside of ancestor subsumption is its cost what led Rolf Socher-Ambrosius to give in \cite{Socher-Ambrosius-Ancestor-subsumption-1992} syntactic characterizations of clause sets giving rise to ancestor subsumption. In \cite{Socher-Ambrosius-Ancestor-subsumption-1992} , such clause sets are said to ``roughly correspond to sets of logical equivalences''. 

Luke Simon, Ajay Bansal, Ajay Mallya and Gopal Gupta proposed in 2006 and 2007 in \cite{Simon-CoLP-2006,Simon-CoLP-2007} \emph{coinductive logic programming}, a form of logic programming which is the first of its kind for two reasons: 
First, it combines the two kinds of codata so far proposed for logic programming, rational (or regular) terms like \verb#X = [0|X]# based on unification without occurs-check, and infinite terms like \verb#f(#$\omega$\verb#)# defined as limits of an infinite sequence  finite terms constructed during the steps of non-terminating SLDF derivations based, according to the greatest fixpoint semantics of logic programs \cite{Abdallah-Metric-Interpretations-and-Greatest-Fixpoint-Semantics-of-LP-1983,Abdallah-Interpretation-of-Infinite-Computations-in-LP-1984,van-Emden-Top-down-semantics-1985,Lloyd-Foundations-of-Logic-Programming-1987}, on unification with occurs-check. 
Second, it introduces into logic programming a proof principle named \emph{coinductive  hypothesis rule} which resembles the coinduction proof principle. Applications of coinductive logic programming are presented in \cite{Gupta-Infinite-Computation-Co-induction-and-Computational-Logic-2011}. Implementations of coinductive logic programming languages are described in \cite{Moura-Portable-Efficient-Implementation-of-Coinductive-LP-2013,Theofrastos-Tabling-Rational-Terms-and-Coinduction-Finally-Together!-2014}. 

The declarative and operational semantics of coinductive logic programming given in \cite{Simon-CoLP-2006,Simon-CoLP-2007} have confusing aspects. One of them is the claim that coinductive logic programming's operational semantics is sound and complete with respect to the greatest fixpoint semantics of logic programs \cite{Simon-CoLP-2006,Simon-CoLP-2007}. No explanations are given of how this relates to the downward approximation sequence $T_p{\downarrow}^n$ of the immediate consequence operator $T_P$, the limit of which specifies the greatest fixpoint semantics of a logic program \cite{van-Emden-Kowalski-Semantics-of-Predicate-Logic-1976}, not reaching a fixpoint before $\omega + 1$ with some (finite and definite) programs $P$  and even not before $\omega_1^{CK}$  with some others  (finite and definite) programs \cite[p. 8]{Fitting-Fixpoint-Semantics-for-Logic-Programming-2002}.\footnote{$\omega$ is the ordinal defined as the set of all finite ordinals and thus the smallest limit ordinal and $\omega_1^{CK}$, the Church-Kleene ordinal, is defined as the set of all recursive ordinals and thus a limit ordinal and the smallest non-recursive ordinal \cite{Church-Kleene-Ordinal-Numbers-1937,Church-Constructive-Second-Number-Class-1938,Kleene-Notations-for-Ordinal-Numbers-1938}.} Another confusing aspect of coinductive logic programming is its ``coinductive  hypothesis rule of the form $\nu(n)$''  \cite[p. 336]{Simon-CoLP-2006}\footnote{The coinductive hypothesis rule and $\nu(n)$ do not seem to be defined in \cite{Simon-CoLP-2006}.} which is defined as follows in \cite[p. 472]{Simon-CoLP-2007}: ``The operational semantics is given in terms of the coinductive hypothesis rule which states that during execution, if the current resolvent $R$ contains a call $C'$ that unifies with a call $C$ encountered earlier, then the call $C'$ succeeds.'' The articles  \cite{Simon-CoLP-2006,Simon-CoLP-2007} do not explain how that coinductive hypothesis rule relates to the coinduction proof principle.
The following examples cast a shadow over the coinductive hypothesis rule:

\begin{example}
\label{ex:P4}
\begin{verbatim} 

   :- coinductive p/2.
   p(X, a) :- p(X, Z).
   p(X, b) :- p(s(X), b).
   
   ?- p(Y, a). 
\end{verbatim}
\end{example}

\begin{example}
\label{ex:P5}
\begin{verbatim} 

   :- coinductive qc/2.
   qc(X, f(X)). 
   
   ?- qc(f(Y), Y).
\end{verbatim}
\end{example}

\begin{example}
\label{ex:P6}
\begin{verbatim} 

   :- inductive q/2.
   qi(X, f(X)), r(X). 
   
   ?- qi(f(Y), Y). 
\end{verbatim}
\end{example}

\begin{example}
\label{ex:P7}
\begin{verbatim} 

   :- inductive pi/1.
   pi(_).
   :- coinductive pc/1.
   pc(_).
   
   ?- X = f(X), pi(X).
   ?- Y = f(Y), pc(Y).
   ?- Z = f(Z), pi(Z), pc(Z).
\end{verbatim}
\end{example}

In  Example \ref{ex:P4}, the SLD derivation from the goal \verb#?- p(Y, a).# is stopped at the subgoal \verb#?- p(Y, Z).# by the coinductive hypothesis rule what prevents the derivation of the subgoal \verb#p(s(Y), b)# and therefore prevents that the rational term \verb#Y = s(Y)# (which expresses $\texttt{s}(\omega)$) be returned as answer. Example \ref{ex:P4} suggests that the coinductive hypothesis rule of coinductive logic programming should not be based on ancestor unification (without occurs-check) but instead on ancestor subsumption (without occurs-check) \cite{Socher-Ambrosius-Ancestor-subsumption-1992}. 

In  Example \ref{ex:P5}, since the predicate \verb#qc# is declared coinductive, the goal \verb#?- qc(f(Y), Y).# is unified (without occurs-check) with the fact \verb#qc(X, f(X)).# yielding as an answer the rational term \verb#Y = f(Y)# even though no clauses in the program specify non-terminating SLD derivations. 

Example \ref{ex:P6} is like Example \ref{ex:P5} except that the predicate, now called \verb#qi# instead of \verb#qc#, is declared inductive instead of coinductive. Since \verb#qc# is declared inductive, a unification with occurs-check of \verb#?- qi(f(Y), Y).# with \verb#qi(X, f(X)).# is attempted which fails because of the occurs-check. 
The articles \cite{Simon-CoLP-2006,Simon-CoLP-2007} give no clues about the intentions behind treating differently cases like Examples \ref{ex:P5} and \ref{ex:P6}.  

In Example \ref{ex:P7}, the goal \verb#?- X = f(X), pi(X).# probably fails because \verb#pi# is declared inductive. Indeed, to the best of the author's understanding, that implies that unifications involving the variable \verb#X# perform the occurs-check. In contrast, the goal \verb#?-Y=f(Y),pc(X).# probably succeeds returning the answer \verb#Y = f(Y)#  because \verb#pi# is declared inductive what, as the author understands, means that unifications involving \verb#Y# do not perform the occurs-check. Note that, according to the examples of \cite[Section 2.5]{Simon-CoLP-2007},  in coinductive logic programming the system predicate \verb#=# is both inductive and coinductive. The three goals of Example \ref{ex:P7} are therefore licit. 
Example \ref{ex:P7} suggests that whether a unification involving a term $t$ performs the occurs-check or not should depend on the type of $t$, data or codata, not on a type inductive or coinductive of predicates in which $t$ occurs. Example \ref{ex:P7} raises the question whether with coinductive logic programming as defined in \cite{Simon-CoLP-2006,Simon-CoLP-2007} a same term $t$ might occur as an argument of both inductive or coinductive predicates, in which case  the third goal \verb#?- Z = f(Z), pi(Z), pc(Z).# is licit but does not seem to be covered by the semantics given in \cite{Simon-CoLP-2006,Simon-CoLP-2007}, or whether inductive and coinductive predicates (except \verb#=#) cannot share arguments, in which case coinductive logic programming as defined in \cite{Simon-CoLP-2006,Simon-CoLP-2007} would be a rather limited approach. 

Coinductive definitions have entered functional programming with lazy evaluation \cite{Henderson-Lazy-evaluation-1976,Friedman-Lazy-evaluation-1976}.  The infinite list of Fibonacci numbers can for example be coded in Haskell as \verb#(f 0 1)# and the finite list of its \verb#43# first elements as \verb#take 43 (f 0 1)# where the function \verb#f# is coded as \verb#f a b = a : f b (a+b)#. 
A type system for coinductively defined data objects, or codata, has been proposed in \cite{CoCaml-2017}. The use of the coinduction proof principle for the analysis of functional programs with codata is discussed in \cite{Gordon-Tutorial-Co-induction-and-FunctionalProgramming-1994}. 
The lazy evaluation of infinite objects has entered logic programming through logic-functional languages \cite{Giovanetti-Kernel-LEAF-1991,Hanus-The-Integration-of-Functions-into-LP}.

Coinductive definitions have been elegantly formalized in 2005 by Jan Rutten as behavioural differential equations \cite{Rutten-A-Coinductive-Calculus-of-Streams-2005}. Rutten's behavioural differential equations are further used in \cite{Kupke-Stream-Differential-Equations-2011,Hansen-Stream-Differential-Equations-2016}. In their term-based form, behavioural differential equations are close to, and extend, the aforementioned representation of rational trees in Prolog II. In the following, coinductive definitions are given in the term-based form of Rutten's behavioural differential equations. 

The coinduction proof principle relates to mathematical induction\footnote{Mathematical induction goes back to the Greek mathematics of the antiquity.} because it relates to structural induction which itself derives from mathematical induction. Structural induction has been formalized in 1959 East of the Iron Curtain by R\'{o}zsa P\'{e}ter in \cite{Peter-Structural-Induction-1961} and in 1969 West of it by Rod M. Burstall \cite{Burstall-Structural-Induction-1969}.\footnote{The idea was, however, already ``in the air'': G{\"o}del's article \cite{Goedel-Unvollstaendigkeitssaetze-1931} on the incompleteness theorems bearing his name among others sketches a proof by mathematical induction on degrees of primitive recursive functions which resembles a proof by structural induction.}

Proof by (mathematical or structural) induction cannot be completely automated because finding a suitable induction hypothesis is neither decidable, nor semi-decidable: ``In contrast to first-order inference, inductive inference is incomplete in the sense that any axiomatization which includes a non-trivial form of induction, there are formulas both true and unprovable'' \cite{Bundy-automation-of-proof-by-mathematical-induction-2006} which follows from Kurt G\"{o}del's first incompleteness theorem \cite{Goedel-Unvollstaendigkeitssaetze-1931}. Proofs by induction have nonetheless be (incompletely) automated \cite{Boyer-Moore-Integrating-decision-procedures-into-heuristic-theorem-provers1988,Bundy-automation-of-proof-by-mathematical-induction-2006,Comon-Inductionless-Induction-2001,Avenhaus-Quod-Libet-2003} in two manners: Explicitly \cite{Bundy-automation-of-proof-by-mathematical-induction-2006} and implicitly after the approach called ``inductive completion'' or ``inductionless induction'' \cite{Comon-Inductionless-Induction-2001}. 
The later approach consists in assuming the conjecture to prove, in attempting to derive an inconsistency using the Knuth-Bendix completion algorithm \cite{Knuth-Bendix-Algorithm-1967}, and if no inconsistency is derived to consider the conjecture proved. Observe that inductive completion (or inductionless induction) resembles the coinduction proof principle.\footnote{This resemblance seems to have so far remained unnoticed.}

Proofs by induction are supported by proof assistants \cite{Geuvers-Proof-Assistants-2009}, that is, proof systems requiring the participation of their human users.
Coinduction as a proof method is usually described as dual of induction, a description referring to the coalgebras of category theory \cite{Jacobs-Rutten-Tutorial-on-CoAlgebras-and-CoInduction-1997}. Like proofs by induction and for the same reason, proofs by coinduction cannot be completely automated. Coinduction as a proof method has been formalized in higher-order logic and automated with the proof assistant Isabelle as reported in \cite{Paulson-Mechanizing-coinduction-and-corecursion-in-higher-order-logic-1997} and in terms of Labelled Transition Systems (LTSs) in \cite{Pous-Sangiorgi-2011}. 
Like proofs by induction, proofs by coinduction have been partly automated in proof assistants \cite{Paulson-Mechanizing-coinduction-and-corecursion-in-higher-order-logic-1997}. 

A proof method which resembles the coinduction proof principle has been proposed for logic programming in 2008 by Joxan Jaffar, Andrew Santosa and R\u{a}zvan Voicu in \cite{Jaffar-Santosa-Voicu-Coinduction-Rule-2008}. The authors stress in this article that the proof method they propose refers to the least fixpoint semantics, not to the greatest fixpoint semantics of of logic programs: ``we would like to clarify that the use of the term coinduction pertains to the way the proof rules are employed for a proof obligation $G \models H$, and has no bearing on the greatest fixed point of the underlying logic program P. In fact, our proof method, when applied successfully, proves that $G$ is a subset of $H$ wrt.\ the least fixpoint of (the operator associated with) the program'' \cite{Jaffar-Santosa-Voicu-Coinduction-Rule-2008}.
In contrast to standard induction and coinduction proof methods, the proof method proposed in \cite{Jaffar-Santosa-Voicu-Coinduction-Rule-2008} can be completely automated: ``a search-based method [\textellipsis] automatically discovers the opportunity of applying induction instead of the user having to specify some induction schema'', ``our method is amenable to automation'' and ``the unfolding process and the application of the coinduction principle require no manual intervention'' \cite{Jaffar-Santosa-Voicu-Coinduction-Rule-2008}.
The generalized coinduction proof principle (see Section \ref{sec:Generalized-Coinduction-Proof-Principle} below) precises how the proof method of \cite{Jaffar-Santosa-Voicu-Coinduction-Rule-2008} relate to the coinduction proof principle. 

Jan Rutten gives in \cite{Jacobs-Rutten-Tutorial-on-CoAlgebras-and-CoInduction-1997,Rutten-universal-coalgebras-2000,Rutten-A-Coinductive-Calculus-of-Streams-2005} coalgebraic treatments of coinduction.

Davide Sangiorgi's book \cite{Sangiorgi-Bisimulation-and-Coinduction-2012} is a comprehensive introduction to coinduction, bisimulation and the coinduction proof principle which, however, does not address coinduction in declarative programming and does not mention inductionless induction \cite{Comon-Inductionless-Induction-2001}. 

\section{Coinductive Definitions}
\label{sec:Coinductive-Definitions}

This section introduces terms of type codata, short \emph{codata terms} after one of the formalisms which have been proposed for the finite representation of constructed streams \cite{Rutten-A-Coinductive-Calculus-of-Streams-2005,Kupke-Stream-Differential-Equations-2011}. Codata terms rely on corecursive definitions which are introduced in the next section. 

Two disjoint data types are considered in the following: \emph{Data} for the representation of non-stream objects and  \emph{codata} for the representation of streams the elements of which are of type data. Considering only the two types data and codata is a convenient simplification. In practice, the type data consists of several pairwise disjoint types like integer, floating-point number, character, etc.\ and structured types like finite list of integers, finite list of floating-point numbers, finite list of characters, etc. Streams of streams could be considered, though, since such streams might make sense for example in specifying stream processing operations but such streams and the issues they raise are not addressed in this article.

The type codata encompasses codata terms and their names. The following illustrates on the Fibonacci function the concept of codata term and of name, base cases  and defining expression of a codata term: 
\begin{quote}
\verb#fib as [0, 1 | fib^0 + fib^1]#
\end{quote}
This expression is a codata term. Its name is \verb#fib#, its base cases are \verb#0# and \verb#1# and its defining expression is \verb#fib^0 + fib^1#. 
In contrast to \cite{Rutten-A-Coinductive-Calculus-of-Streams-2005}, the keyword \verb#as# is used instead of \verb#=# for avoiding a confusion with the the Prolog fashion's use of \verb#=# in the former section of this article for denoting unification. Codata terms as illustrated above and defined below slightly generalise those of \cite{Rutten-A-Coinductive-Calculus-of-Streams-2005} by allowing compound codata names which are needed for expressing mutual corecursion. 

It is convenient to think of the codata term \verb#fib# as a definition of an infinite list, a view which is justified below in Section \ref{sec:Interpretation-of-Codata-as-Infinite-Streams}. \verb#fib^0# denotes \verb#fib# and for all natural number \verb#n#, \verb#fib^(n+1)# denotes the tail of \verb#fib^(n)#. Thus, \verb#fib^1# denotes the tail of \verb#fib#. 

If \verb#f# and \verb#g# respectively represent  
\begin{quote}
\verb#[f(0), f(1), ..., f(n), ...]# \\
\indent
\verb#[g(0), g(1), ..., g(n), ...]#
\end{quote}
then \verb#f+g# represents
\begin{quote}
\verb#[f(0)+g(0), f(1)+g(1), ..., f(n)+g(n), ...]# 
\end{quote}
(see below Section \ref{sec:Interpretation-of-Codata-as-Infinite-Streams}). 
Thus, the definition
\begin{quote}
 \verb#fib as [0, 1 | fib^0 + fib^1]# 
 \end{quote}
can be step-wise expanded yielding 
\begin{quote}
\verb#[0, 1 | fib^0 + fib^1]# \\
\indent
\verb#[0, 1, 1 | fib^1 + fib^2]# \\
\indent
\verb#[0, 1, 1, 2 | fib^2 + fib^3]# \\
\indent
\verb#[0, 1, 1, 2, 3 | fib^3 + fib^4]# \\
\indent
\verb#[0, 1, 1, 2, 3, 5 | fib^4 + fib^5]# \\
\indent
etc.
\end{quote}

Even though the codata term \verb#fib# is a recursive definition of the Fibonacci function, it specifies an iterative computation, as does a tail recursive definition \cite{Friedman-Wise-Tail-Recursion-1974,Steele-Tail-Recursion-1977}, but it does not require accumulators, in contrast to a tail recursive definition. This is an essential aspect of coinductive definitions which can be expressed in logic programming parlance as follows: Codata are meant to express forward chaining inferences as defined by the immediate consequence operator $T$ \cite{van-Emden-Kowalski-Semantics-of-Predicate-Logic-1976}. 

In the following, Data denotes the set of terms of type data and Codata denotes the set of terms of type codata. 
In Section \ref{sec:Interpretation-of-Codata-as-Infinite-Streams} below, it is shown that a codata term \verb#s as [s#$_0$\verb#, s#$_1$\verb#, ..., s#$_n$\verb# | E]#  with \verb#s# a constant induces the definition of a function: 
\begin{quote}
\verb#s#: $\mathbb{N} \rightarrow$ Data
\end{quote}
This justifies to view \verb#s# as a definition of the infinite list 
\begin{quote}
\verb#[s(0), s(1), s(2),..., s(n), ...]# 
\end{quote}
and the following definitions. 

\begin{definition}[Codata names]
A \emph{codata name} is either a codata constant (called an \emph{atomic codata name}), or a tuple of codata constants (called a \emph{compound codata name)}. 

If $(\texttt{s}_1, \ldots \texttt{s}_m)$ with $m \geq 2$ is a compound codata name and if $n \in \mathbb{N} \setminus \{0\}$, then $(\texttt{s}_1, \ldots \texttt{s}_m)(n)$ denotes $(\texttt{s}_1(n), \ldots \texttt{s}_m(n))$. 
\end{definition}

\begin{definition}[Head, tail, $n$-tail, $n$th element of a codata term]
\label{def:head-tail}
Let  \verb#s# be a codata name. 
\begin{itemize}
   \item  \verb#s(n)# denotes the \emph{$n$th element} of \verb#s#.  
   \item \verb#s(0)# is the \emph{head} of \verb#s#.  
   \item \verb#s^0# = \verb#s#
   \item \verb#s^1# is the \emph{tail} of \verb#s# which expresses the infinite list \\
            \verb#[s(1), s(2),..., s(n), ...]#
   \item for all $n \in \mathbb{N}$ \verb#s^(n+1)# = \verb#s^n^1# 
   \item for all $n \in \mathbb{N}$ \verb#s^n# is the \emph{$n$th-tail} of \verb#s# 
\end{itemize}

\end{definition}
It follows from Definition \ref{def:head-tail} by mathematical induction that for all $n \in \mathbb{N}$ and $m \in \mathbb{N}$ \verb#s^n^m# = \verb#s^(n+m)#.

\begin{definition}[Codata term]
A \emph{codata term} or \emph{codata definition} is an expression of the form \verb#s as [s#$_0$\verb#, s#$_1$\verb#, ..., s#$_n$\verb# | E]# where:
\begin{itemize}
   \item \verb#s# is an atomic or compound codata name.
   \item $n \geq 0$.
   \item the \verb#s#$_i$, $0 \leq i \leq n$, are terms of sort data called the \emph{base cases}.
   \item \verb#E# is a \emph{codata expression}.
\end{itemize}
\verb#s as [s#$_0$\verb#, s#$_1$\verb#, ..., s#$_n$\verb# | E]#  is a \emph{definition} of \verb#s# and \verb#s# is called the \emph{name} of the codata term. 
\end{definition}

Codata expressions are defined below in Section \ref{sec:Corecursive-Definitions}. 

\begin{example}
The  following mutually recursive functions 
\begin{quote}
\begin{tabular}{l c l l}
$f(0)$ 	& $=$ 	&$0$ \\
$g(0)$ 	& $=$ 	& $1$ \\
$f(n)$	& $=$	& $g(n-1)$ 		& for $n \geq 1$ \\
$g(n)$	& $=$	& $2 \times f(n-1)$ 	& for $n \geq 1$  \\
\end{tabular}
\end{quote}
can be expressed by the following codata term: 
\begin{quote}
\begin{verbatim}
(f, g) as [(0, 1) | (g, 2*f)]
\end{verbatim}
\end{quote}
\end{example}

Defining expressions like \verb#2*f# are defined below in Section \ref{sec:Corecursive-Definitions}. 

\begin{definition}[Well-formed codata term]
\label{def:Well-formed-codata-term}
A codata term 
\begin{quote}
\verb#s as [s#$_0$\verb#, s#$_1$\verb#, ..., s#$_n$\verb# | E]# 
\end{quote}
is \emph{well-formed} if 
\begin{enumerate}
   \item For all $i = 1, \ldots, n$, \verb#s#$_i$  and \verb#s# are either both atomic or tuples of the same size.
   \item If \verb#s# is atomic, then \verb#E# defines an atomic codata, else \verb#E# defines a tuple of the size of  \verb#s#. 
   \item All atomic names occurring in the defining expression \verb#E# also occur in the codata name \verb#s#.
   \item If the defining expression \verb#E# refers to a $k$th tail \verb#t^k#, then $n \geq k+1$. 
\end{enumerate}
\end{definition}
The first two conditions ensure that the name, the base cases and the defining expression of a codata term are consistent in their sizes. The third conditions ensures that every name referred to in a codata term is defined in that codata term. The fourth and last condition ensures that the number of base cases in a codata term and the references to tails in that codata term's defining expression are consistent.  

\begin{example}
Using \verb#nat#, a codata term expressing the sequence of natural numbers, Douglas Hofstadter's female and male sequences \cite{Hofstadter-Goedel-Escher-Bach-1979}:
\begin{quote}
\begin{tabular}{l c l l}
$f(0)$ 	& $=$ 	&$1$ \\
$m(0)$ 	& $=$ 	& $0$ \\
$f(n)$	& $=$	& $n - m(f(n-1))$ 		& for $n \geq 1$ \\
$m(n)$	& $=$	& $n - f(m(n-1))$ 	& for $n \geq 1$  \\
\end{tabular}
\end{quote}
can be expressed by the following codata term: 
\begin{quote}
\begin{verbatim}
(nat,f,m) as [(0,1,0) | (1+nat,nat-m^f,nat-f^m)]
\end{verbatim}
\end{quote}
\end{example}
This example shows that codata terms can express \emph{non-linear} recurrence relations. Such a definition requires a more involved notion of well-formedness than that given above in Definition \ref{def:Well-formed-codata-term}, though. This more involved notion of well-formedness is out of the scope of this article. 

\section{Corecursive Definitions}
 \label{sec:Corecursive-Definitions}
 
The qualifier ``inductive'' (``coinductive'', respectively) refers to data (codata, respectively) definitions while the qualifier ``recursive'' (``corecursive'', respectively) refers to the definitions of functions or predicates on data (codata, respectively). In most cases, the pedantic distinction (co)inductive/(co)recursive can be ignored. Furthermore, this distinction is blurred in the presence of codata which can be seen both as infinite lists and function definitions. The distinction is nonetheless used in the titles of this section and of the former section because it provides a convenient structuring of the exposition.  

In the following, \emph{pointwise} means element-wise and refers to the interpretation of codata terms as infinite lists which is justified below in Section \ref{sec:Interpretation-of-Codata-as-Infinite-Streams}. 
In the following, all corecursive definitions refer to natural numbers. Example of corecursive referring to other data types (like Boolean, characters, strings of characters, etc.) can easily be derived from the following ``number-based'' examples. 

If \verb#a#$_d$ is a data constant, then the \emph{codata constant} \verb#a#$_c$ is defined by: \\
\verb#a#$_c$\verb# as [a#$_d$ \verb#| a#$_c$\verb#]#

The \emph{sum} of a data constant \verb#a#$_d$ and a codata \verb#s# is defined by:  \\
\verb#a#$_d$\verb#+s as [a#$_d$\verb# + s(0) | a#$_d$\verb# + s^1]#

The \emph{pointwise product} of codata \verb#s# and \verb#t# is defined by: \\
\verb#s*t as [s(0) * t(0) | s^1 * t^1]#

The codata \verb#nat# (natural numbers) is defined by: \\ 
\verb#nat as [0 | 1#$_d$\verb# + nat]#. 

The \emph{pointwise sum} \verb#s+t# of codata \verb#s# and \verb#t# is defined by: \\
\verb#s+t as [s(0) + t(0) | s^1 + t^1]#

The codata \verb#nat# (natural numbers) can also be defined by: \\
\verb#nat as [0 | 1#$_c$\verb# + nat]#

The \emph{product} of a data constant \verb#a#$_d$ and a codata \verb#s# is defined by: \\
\verb#a#$_d$\verb#*s as [a#$_d$\verb# * s(0) | a#$_d$\verb# * s^1]#

The \emph{pointwise pairing} of codata  \verb#s# and \verb#t# is defined by: \\
\verb#(s, t) as [(s(0), t(0)) | (s^1, t^1)]#

The \emph{pointwise application of data term constructor} $f$ to codata  \verb#s# is defined by: \\
\verb#f(s) as [f(s(0)) | f(s^1)]#

Right-side sums and products with data constants are defined similarly to the left-side sums and products with data constants specified above. Pointwise $n$-groupings for $n \geq 3$ of codata are defined similarly to the pairing of codata. 

%
%
Further examples of corecursive definitions \cite{Rutten-A-Coinductive-Calculus-of-Streams-2005}, where \verb#s# and \verb#t# are codata names:
\noindent
\begin{verbatim}
even(s) as [s(0) | even(s^2)]
odd(s) as [s^1(0) | odd(s^2)]
split(s) as (even(s), odd(s))
zip(s, t) as [s(0) | zip(t, s^1)]
\end{verbatim}

The codata \verb#fact# (factorial numbers) is defined by: 
\noindent
\begin{verbatim}
fact as [1 | nat^1 * fact]
\end{verbatim}

%

\begin{definition}[Codata expression]
\label{def:Codata-expression}
A \emph{codata expression} is a codata $n$-th tail (including a codata name) or the application of a corecursive function to codata expression(s).

Let 
\verb#s as [s#$_0$\verb#, s#$_1$\verb#, ..., s#$_n$\verb# | E]# 
be a codata term with an atomic name. \verb#s#  \emph{immediately depends} on every codata constant and every codata name occurring in the codata expression \verb#E#. \emph{Depends} is used for the transitive closure of \emph{directly depends}. 
\end{definition}

Immediate dependency extends component-wise to compound codata names.  

\begin{definition}[Structural codata expression]
\label{def:Structural-codata-expression}
A codata expression is \emph{structural} if it contains neither sums, nor products. 
\end{definition}
More generally, a function definition is structural if its does not involve expressions determining values of which requires reductions (that is, computations). 

Observe that the corecursive function definitions given above are all pointwise. 
Observe also that the codata constants, the codata pairing and grouping and the pointwise data term application are the only structural cases of structural corecursion given above. 

\section{Interpretation of Codata as Infinite Streams}
\label{sec:Interpretation-of-Codata-as-Infinite-Streams}

This section shows that a codata term \verb#s as [s#$_0$\verb#, s#$_1$\verb#, ..., s#$_n$\verb# | E]#  specifies an infinite list in the sense that it induces the definition of a function \verb#s#: $\mathbb{N}^m \rightarrow$ Data where $m$ is the dimension of \verb#s#, that is, $m = 1$ if \verb#s# is atomic, and $m$ is the tuple size of \verb#s# otherwise. The overloading of the symbol \verb#s# used both as a codata name and a function name is intentional and justified by Definition \ref{def:head-tail} of Section \ref{sec:Coinductive-Definitions}. 

\begin{definition}
\label{def:tailed-expressions}
Let \verb#s as [d#$_1$\verb#, ... , d#$_k$\verb# | E]# be a codata term. 

The notation \verb#E[s#$_0$\verb#, ..., s#$_n$\verb#]# expresses that the set of atomic codata names occurring in expression \verb#E# is $\{$\verb#s#$_0, \ldots,$ \verb#s#$_n\}$, that is, for all $0 \leq i \leq n$, \verb#s#$_i$ $=$ \verb#s(i)#. 

\verb#E^0# denotes \verb#E# and if $k \in \mathbb{N} \setminus \{0\}$, then \verb#E^k# denotes the expression obtained from expression \verb#E# by replacing every $n$th-tail \verb#t^n# in \verb#E# by \verb#t^n^k#.
\end{definition}
Thus, by Definition \ref{def:tailed-expressions}, if $k \in \mathbb{N}$, then 
\begin{quote}
\verb#E[s#$_0$\verb#, ..., s#$_i$\verb#, ..., s#$_n$\verb#]^k# 
\end{quote}
denotes  
\begin{quote}
\verb#E[s#$_0$\verb#^k, ..., s#$_i$\verb#^k, ..., s#$_n$\verb#^k]#.
\end{quote}

\begin{definition}[$n$th expansion of a codata term]
\label{def:nth-expansion}
Let 
\begin{quote}
\verb#s as [s#$_0$\verb#, s#$_1$\verb#, ..., s#$_k$\verb# | E]# 
\end{quote}
be a codata term and let $n \geq k$ be a natural number. The \emph{$n$th expansion} of \verb#s# is the codata term 
\begin{quote}
\verb#s[n] as [s(0), ..., s(k), ..., s(n) | E^(n-k)]# 
\end{quote}
obtained from \verb#s# by repeatedly applying the codata term defining expression \verb#E#. 
Recall that for all $0 \leq i \leq k$, \verb#s#$_i$ $=$ \verb#s(i)#. 
\end{definition}
Section \ref{sec:Coinductive-Definitions} above gives examples of expansions of the codata term \verb#fib#. 
The following shows that the applications mentioned in Definition \ref{def:nth-expansion} terminate if the codata term considered is well-formed.

\begin{proposition}[Expansion Theorem]
\label{prop:Expansion-Theorem}
For all well-formed codata terms \verb#s as [s#$_0$\verb#, s#$_1$\verb#, ..., s#$_k$\verb# | E]# and all $n \in \mathbb{N}$ such that $n \geq k$, the $n$th expansion \verb#s[n]# of \verb#s# is uniquely defined. 
\end{proposition}

\begin{proof}
First observe that if \verb#s as [s#$_0$\verb#, s#$_1$\verb#, ..., s#$_k$\verb# | E]#  is a well-formed codata term, then \verb#s# (directly or indirectly) depends on finitely many codata names all occurring in \verb#s#. Second, observe that \verb#[s#$_0$\verb#, s#$_1$\verb#, ..., s#$_k$\verb# | E]# is its $k$th expansion of \verb#s#.
The result follows by  structural induction. 
\end{proof}

\begin{definition}[$n$th element of a codata term]
\label{def:nth-element}
Let 
\begin{quote}
\verb#s as [s#$_0$\verb#, s#$_1$\verb#, ..., s#$_n$\verb# | E]# 
\end{quote}
be a well-formed codata term and let $m \in \mathbb{N} \setminus \{0\}$ be the dimension of \verb#s#. The codata term \verb#s# induces a function 
\begin{quote}
\verb#s#: $\mathbb{N}^m \rightarrow$ Data 
\end{quote}
defined by: for $n \in \mathbb{N}$ \verb#s(n)# is the head of the $n$th expansion of \verb#s#. 
\end{definition}
By the Expansion Theorem, the function \verb#s#: $\mathbb{N}^m \rightarrow$ Data referred to in Definition \ref{def:nth-element} is well-defined. 
As a consequence, a well-formed codata term \verb#s# can be viewed as a definition of the infinite list 
\begin{quote}
\verb#[s(0), s(1), s(2),..., s(n), ...]# 
\end{quote}
and the concepts of Definition \ref{def:head-tail} from Section \ref{sec:Coinductive-Definitions} above are well-defined. 

\begin{corollary}[tail computation]
\label{cor:nth-tail-computation}
Let $\texttt{s} = (\texttt{s}_0, \texttt{s}_1, \ldots, \texttt{s}_k)$ with $k \geq 1$ be the name of a well-formed codata term.
\begin{enumerate}
   \item For all $n \in \mathbb{N}$ and $m \in \mathbb{N}~ s{\myHat}n{\myHat}m =$ \verb#s#${\myHat}(n+m) =$ \verb#s#${\myHat}n(m) =$ \verb#s#${\myHat}m(n)$
   \item If \verb#s as [d#$_1$\verb#, ..., d#$_k$\verb# | E[s#$_0$\verb#, ..., s#$_k$\verb#)]]#, then for all $n \in \mathbb{N}$, if $n \geq k$ then \\
             \verb#s^n = E[s#$_0$\verb#^(n-k), ..., s#$_k$\verb#^(n-k)]#. 
\end{enumerate}
\end{corollary}

\begin{proof}
By structural induction and the Expansion Theorem \ref{prop:Expansion-Theorem}.
\end{proof}

A few remarks are worthwhile. 

First, even though the set of infinite lists on a finite alphabet is not enumerable,\footnote{This can be proven by the diagonalisation argument Georg Cantor used for proving that the set of decimal numbers is not countable \cite{Cantor-Diagonalization-1891}.} the set of codata terms on a finite alphabet is recursively enumerable. Indeed, like a data term, codata terms are finite expression which can be \emph{inductively} defined. Thus, using the terminology of the introduction, not all observed streams can be specified as constructed streams.

Second, two distinct codata terms might specify the same infinite stream as is the case for example with \verb#s1 = [1 | s1]# and \verb#s2 = [1, 1 | s2]# which both specify the same codata constant specifying the infinite list of $1s$: \verb#[1, 1, 1, ...]#.

Third, let \verb#s# and \verb#t# be the names of two well-formed codata terms. Whether for all $n \in \mathbb{N}$ \verb#s(n) = t(n)# (that is, \verb#s# and \verb#t #specify the same infinite list) is neither decidable nor semi-decidable. Algorithms are possible, though, which in some cases will decide whether, or recognize that, two codata terms specify a same infinite list or whether all elements of the infinite list specified by a codata term has a certain property. 
 
Fourth and finally, it is worth stressing that one of the interests of codata terms is to (corecursively) define functions $f: \mathbb{N} \rightarrow$ Data without having to give an equation on $n$ for $f(n)$. There are indeed cases in which this is desirable. The Fibonacci function for example has a simple recursive definition,  \verb#fib as [0, 1 | fib + fib^1]#, but a significantly more complicated definition as an equation on $n$: 
\begin{quote}
$\textrm{fib}(n) = \round{\frac{(1 + \sqrt{5})^n}{2^n \sqrt{5}}}$
\end{quote}
In other cases, however, an equation on $n$ is simpler than a recursive definition: For all $n \in \mathbb{N}~ f(n) = n$ is for example simpler than: 
\begin{quote}
\verb#nat as [0 | 1 + nat]#
\end{quote}

\section{First Generalized Coinduction Theorem}

In this section and the following, a predicate is called a \emph{data predicate} if all its arguments are of type data and a predicate is called a \emph{codata predicate} if all its arguments are of type codata, 
data predicates are denoted by identifiers beginning with lower case characters and codata predicates are denoted by identifiers beginning with upper case characters, codata terms are implicitly assumed to be well-formed and the phrase ``the codata term \verb#s#'' refers to a (well-formed) codata term of the form \verb#s as [d#$_1$\verb#, ... , d#$_k$\verb# | E]# with $k \geq 1$. 

\begin{definition}[Pointwise-defined and hereditary codata predicate]
\label{def:Pointwise-defined-hereditary}
Let $p$ be a unary data predicate and $s$ a codata name. 

The unary codata predicate $P_p$ defined by:
\begin{quote}
 For all codata term $s$, $P_p(s)$ iff for all $n \in \mathbb{N}~ p(s(n))$
 \end{quote}
 is called the \emph{pointwise-derived} from $p$.
 
A unary codata predicate $H$ which fulfills the following condition is said to be \emph{hereditary with respect to} $p$, short \emph{$p$-hereditary}: \\
If $H(s)$ holds of some codata $s$, then
\begin{enumerate}
   \item $p(s(0))$ holds.
   \item $H(s{\myHat}1)$ holds.
\end{enumerate}
\end{definition}

\begin{lemma}
\label{lemma:pointwise-is-hereditary}
Let $p$ be a unary data predicate. The pointwise-derived $P_p$ from $p$ is $p$-hereditary. 
\end{lemma}

\begin{proof}
Let $p$ be a unary data predicate and let $s$ be a codata name. By definition of the pointwise-derived $P_p$ from $p$,  $P_p(s)$ iff for all $n \in \mathbb{N}~ p(s(n))$. Thus, if $P_p(s)$ holds, then $p(s(0))$ holds and $P_p(s{\myHat}1)$ holds.
\end{proof}

Observe that the concept of $p$-hereditary codata predicate generalises that of bisimulation relation \cite{Park-Concurrency-on-automata-and-infinite-sequences-1981,Rutten-A-Coinductive-Calculus-of-Streams-2005,Sangiorgi-Bisimulation-and-Coinduction-2012} since a bisimulation relation $B$ is defined as a binary relation such that $s_1 ~B~ s_2$ if 
\begin{enumerate}
   \item $s_1(0) = s_2(0)$
   \item $s_1{\myHat}1 ~B~ s_2{\myHat}1$
\end{enumerate}
More precisely, a bisimulation relation $B$ is an =-hereditary predicate applying of pair codata $(s_1, s_2)$. Thanks to codata pairing defined above in Section  \ref{sec:Corecursive-Definitions}, a binary codata predicate can be seen as a unary codata predicate the argument of which is a pair of codata.

Observe also that if $p$ is a unary data predicate and $P_1$ and $P_2$ are $p$-hereditary codata predicates, then the codata predicates $A$ and $O$ defined as follows are also $p$-hereditary:
\begin{quote}
For all codata names $s$:
\begin{itemize}
   \item $A(s)$ iff $(P_1(s) \land P_2(s))$
   \item $O(s)$ iff $(P_1(s) \lor P_2(s))$
\end{itemize}
\end{quote}

\begin{definition}[Maximal $p$-hereditary codata predicate]
\label{def:max-p-hereditary-codata-predicate}
Let $p$ be a unary data predicate.
The \emph{maximal p-hereditary codata predicate} $\textrm{Max}_p$ is defined by:
\begin{quote}
If $H$ is a $p$-hereditary codata predicate and s is as codata name such that $H(s)$ holds, then $\textrm{Max}_p(s)$ holds. 
\end{quote}
\end{definition}

Observe that  $\textrm{Max}_p$ is the greatest $p$-hereditary predicate in the sense that it expresses the union of all unary relations on codata which are defined by unary $p$-hereditary codata predicates. 
Observe also that $\textrm{Max}_p$ generalizes the bisimilarity relation \cite{Park-Concurrency-on-automata-and-infinite-sequences-1981,Rutten-A-Coinductive-Calculus-of-Streams-2005,Sangiorgi-Bisimulation-and-Coinduction-2012} which, in the terminology of Definition \ref{def:max-p-hereditary-codata-predicate}, is $\textrm{Max}_=$. 

\begin{theorem}[First Generalized Coinduction Theorem]
\label{thm:First-Generalized-Coinduction-Theorem}
Let $s$ be the name of a codata term and let $p$ a unary data predicate.
The following statements are equivalent: 
\begin{enumerate}
   \item There is a unary $p$-hereditary codata  predicate $H$ such that $H(s)$ holds. 
   \item There is a unary $p$-hereditary codata predicate $H$ such that: $\forall n \in \mathbb{N}~ H(s{\myHat}n)$
   \item $\forall n \in \mathbb{N}~ p(s(n))$
\end{enumerate}
\end{theorem}

\begin{proof}
Let $s$ be a codata term, $p$ a unary data predicate and $H$ a unary codata predicate. 

(1) $\Rightarrow$ (2): 
Assume $H$ is $p$-hereditary and $H(s)$ holds. The proof is by induction. Since by hypothesis H is $p$-hereditary and $H(s)$ holds, $H(s{\myHat}1)$ holds.
Let $n \in \mathbb{N}$ and assume that $H(s{\myHat}n)$ holds.  Since by hypothesis H is $p$-hereditary and $H(s{\myHat}n)$ holds, $H(s{\myHat}(n+1))$ holds.

(2) $\Rightarrow$ (3): 
Since by assumption for all $n \in \mathbb{N}~ H(s{\myHat}n)$ holds and $H$ is $p$-hereditary, for all $n \in \mathbb{N}~ p(s{\myHat}n(0))= p(s(n))$ holds. 

(3) $\Rightarrow$ (1): 
If for all $n \in \mathbb{N}~ p(s(n))$, then $P_p(s)$ holds where $P_p$ is the pointwise-derived from $p$ defined in Definition \ref{def:Pointwise-defined-hereditary}. By Lemma \ref{lemma:pointwise-is-hereditary}, $P_p$ is $p$-hereditary.  
\end{proof}

\section{Generalized Coinduction Proof Principle}
\label{sec:Generalized-Coinduction-Proof-Principle}

Recall that all codata terms considered are implicitly  assumed to be well-formed. 

\begin{corollary}[Generalized Coinduction Proof Principle]
\label{cor:Generalized-Coinduction-Proof-Principle}
Let $p$ be a unary data predicate and let $s$ be the name of a codata term. 
For proving $~ \forall n \in \mathbb{N}~ p(s(n))~$
\begin{enumerate}
   \item it suffices to exhibit a unary codata predicate $H$ such that: 
             \begin{enumerate} 
                 \item $H(s)$ holds. 
                 \item $H$ is $p$-hereditary. \\
                          (That is, for all codata names $t$, $p(t(0))$ and $H(t{\myHat}1)$ holds.)
             \end{enumerate}
   \item it suffices to exhibit a unary $p$-hereditary codata predicate $H$ such that $\forall n \in \mathbb{N}~ H(s{\myHat}n)$.
\end{enumerate}
\end{corollary}

\begin{proof}
(1) is (1) $\Rightarrow$ (3) of Theorem \ref{thm:First-Generalized-Coinduction-Theorem} and 
(2) is (2) $\Rightarrow$ (3) of Theorem \ref{thm:First-Generalized-Coinduction-Theorem}.
\end{proof}

The generalized coinduction proof principle extends to every unary data predicate the coinduction proof principle formulated in \cite{Park-Concurrency-on-automata-and-infinite-sequences-1981,Rutten-A-Coinductive-Calculus-of-Streams-2005,Sangiorgi-Bisimulation-and-Coinduction-2012} for equality. 

If several codata terms mutually depend on each other (in the sense of Definition \ref{def:Codata-expression}, that is, are \emph{mutually corecursive}), then a proof by strong coinduction referring to one of these codata names $s$ must, of course, also refer to the codata names $s$ depends on. This stresses the necessity for the codata terms considered to be well-formed.

For a better understanding, the following examples of proofs by (generalized) coinduction are detailed (and possibly over-detailed). 

\begin{example}
\label{ex:1>0}
\begin{quote}
\verb#1#$_c$\verb# as [1 | 1#$_c$\verb#]#
\end{quote}
The singleton set consisting of this codata term is well-defined. Let $P$ be the codata predicate defined by: For all codata name $s$, $P(s)$ if $s(0) > 0$ and $s{\myHat}1 = s$. Thus, $P$ is $(>0)$-hereditary and $P(\texttt{1}_c)$ holds. By the generalized coinduction proof principle, $\forall n \in \mathbb{N}~$ \verb#1#$_c(n) > 0$. 
\end{example}

\begin{example}
\label{ex:fib>0}
\begin{quote}
\verb#fib = [0, 1 | fib + fib^1]#
\end{quote}
The singleton set consisting of this co-data term is well-defined. 
Proof by generalized coinduction of $\forall n \in \mathbb{N}~ (n > 0 \Rightarrow$ \verb#fib#$(n) > 0)$. 

Let $P$ be the codata predicate defined by: For all codata names $s$, $P(s)$ holds if $s(0) > 0$ and $\forall n \in \mathbb{N}~ ((s(n) > 0 \land s(n+1) > 0) \Rightarrow s(n+2) > 0)$. 

First, $P$ is shown to be $(>0)$-hereditary. Let $s$ be a codata name. If $P(s)$ holds, then by definition of $P$, $s(0) >0$ and $\forall n \in \mathbb{N}~ ((s(n) > 0 \land s(n+1) > 0) \Rightarrow s(n+2) > 0)$. As a consequence, $s(1) >0$ and  $\forall n \in \mathbb{N} \setminus \{0\}~ ((s(n) > 0 \land s(n+1) > 0) \Rightarrow s(n+2) > 0)$, that is, $P(s{\myHat}1)$ holds. 

Second, it is proved that $P($\verb#fib^1#$)$ holds. 
\verb#fib^1#$(0) =$ \verb#fib#$(1) = 1 > 0$ and \verb#fib^1#$(1)$ = \verb#fib#$(2) = 2 > 0$. 
Let $n \in \mathbb{N}$. 
From
\begin{quote}
\verb#fib = [0, 1 | fib + fib^1]#
\end{quote}
it follows by Corollary \ref{cor:nth-tail-computation} that
\begin{quote}
\verb#fib^1^2 = fib^1 + fib^1^1#
\end{quote}
Thus, for all \verb#n#  $\in \mathbb{N}$: 
\vspace{-2em}
\begin{quote}
\noindent
\begin{verbatim}
fib^1(n+2) = fib^2(n+1) = 
fib(n+1) + fib^1(n+1) = 
fib^1(n) + fib^1(n+1)
\end{verbatim}
\end{quote}
It follows that if \verb#fib^1(n)# $> 0$ and \verb#fib^1(n+1)# $> 0$, then \\
\verb#fib^1(n+2)# $> 0$. 
Thus $P($\verb#fib#${\myHat}1)$ holds. 

The result follows by the generalized coinduction principle.
\end{example}

\begin{example} 
\label{ex:zip(even(s),odd(s))=s}
This example is adapted from \cite{Rutten-A-Coinductive-Calculus-of-Streams-2005}.
\vspace{-2em}
\begin{quote}
\noindent
\begin{verbatim}
even(s) as [s(0)|even(s^2)]
odd(s) as [s^1(0)|odd(s^2)]
zip(s, t) as [s(0)|zip(t, s^1)]
\end{verbatim}
\end{quote}
This set of codata terms is well-defined. 
Proof by coinduction for all codata name \verb#s# and \verb#t# \\
\verb#even(zip(s, t)) = s#.

Let $P$ be the binary codata defined by: \\
$(\star)$ for all codata names \verb#s# and \verb#t#, $P$\verb#(even(zip(s, t)), s)#.

$P$ is =-hereditary, that is, if $P(u, v)$ holds, then 
\begin{enumerate}
   \item $u(0) = v(0)$
   \item $P(u{\myHat}1, v{\myHat}1)$
\end{enumerate}

(1) follows from the fact that $(\star)$ implies
\verb#even(zip(s, t))(0) =# \verb#s(0)# 
what is proven as follows: 

\medskip 

\begin{tabular}{l l}
\verb#even(zip(s, t))(0) =# 	& (by def.\  of \verb#even#) \\ 
\verb#zip(s, t)(0) =#			& (by def.\ of \verb#zip#) \\
\verb#s(0)#
\end{tabular}

\medskip 

(2) follows from the fact that $(\star)$ implies that for all codata names \verb#s# and \verb#t#\\ 
$P$\verb#(even(zip(s, t))^1, s^1)# holds, as it is now shown. 
Let \verb#s# and \verb#t# be codata names. 
\begin{verbatim}
even(zip(s, t))^1 = 
even(zip(s, t)^2) = 
even(zip(s, t)^1) = 
even(zip(t, s^1)^1) = 
even(zip(s^1, t^1))
\end{verbatim}
Thus, $P$\verb#(even(zip(s, t))^1, s^1)# holds iff:
\begin{quote}
$P$\verb#(even(zip(s^1, t^1), s^1)# 
\end{quote}
holds. By $(\star)$, 
\begin{quote}
$P$\verb#(even(zip(s^1, t^1), s^1)#
\end{quote}
holds, that is, 
\begin{quote}
$P$\verb#(even(zip(s, t))^1, s^1)# 
\end{quote}
holds. 
By the coinduction proof principle, it follows that:
\begin{quote}
 \verb#even(zip(s, t)) = s#.
 \end{quote}
\end{example}

The proof by coinduction of \verb#zip(even(s), odd(s)) = s# of Example \ref{ex:zip(even(s),odd(s))=s} given above stresses an aspect of some proofs by coinduction. A codata predicate $P$ is constructed as the closure over tails of codata,  what is often described as the computation of a greatest fixpoint \cite{Rutten-A-Coinductive-Calculus-of-Streams-2005,Sangiorgi-Bisimulation-and-Coinduction-2012}. This view is recalled in the following. 

Let $p$ be a unary data predicate and consider the following operator $\Phi_p$ on the unary codata relations: 
\begin{quote}
\begin{tabular}{l c l c}
 $\Phi_p$: 	& $\mathcal{P}(\mathrm{Codata})$	& $\rightarrow$ 	& $\mathcal{P}(\mathrm{Codata})$ \\

			& $R$						&  $\rightarrow$ 	& $\{s \in \mathrm{Codata} \mid p(s(0)) \land R(s{\myHat}1)\}$
\end{tabular}
\end{quote}
If $s$ is a codata name, a proof by generalized coinduction of
\begin{quote}
$\forall n \in \mathbb{N}~ p(n)$ 
\end{quote}
consists in establishing the existence of a unary $p$-hereditary predicate $R$, which defines a unary relation on $\mathcal{P}(\mathrm{Codata})$, such that $R(s)$ holds, that is, that $R \subseteq \Phi_p(R)$ that means that $R$ is a post-fixpoint of $\Phi_p$. By a theorem of Knaster \cite{Knaster-1928,Tarski-1930,Tarski-1955} commonly called the Knaster-Tarski theorem, the greatest fixpoint of $\Phi_p$ is the unary relation on Codata characterized by the predicate $\textrm{Max}_p$ defined above in Definition \ref{def:max-p-hereditary-codata-predicate}.\footnote{The greatest fixpoint $\textrm{Max}_p$ of $\Phi_p$ is unrelated to the greatest fixpoint semantics of logic programs \cite{Abdallah-Metric-Interpretations-and-Greatest-Fixpoint-Semantics-of-LP-1983,Abdallah-Interpretation-of-Infinite-Computations-in-LP-1984,van-Emden-Top-down-semantics-1985,Lloyd-Foundations-of-Logic-Programming-1987} mentioned in Section \ref{set:Coinduction-in-Programming-so-far}.}

The need for a strengthening of the definition of the predicate to be proven $p$-hereditary in a proof by coinduction, as in the proof given in Example \ref{ex:zip(even(s),odd(s))=s} above, is not specific of coinduction proofs. Such strengthening are common in induction proofs where they are called ``induction loading''. A traditional example of induction loading is that a proof by induction of 
\begin{quote}
$$\forall n \in \mathbb{N} \setminus \{0\}~  \sum_{i=1}^{i=n} \frac{1}{i^2} < 2$$ 
\end{quote}
which is impossible without strengthening that statement for example into: 
\begin{quote}
$$\forall n \in \mathbb{N} \setminus \{0\}~  \sum_{i=1}^{i=n} \frac{1}{i^2} \leq 2 - \frac{1}{n}$$
\end{quote}

\begin{example}  
\label{ex:sum}
Proof by coinduction of:
\begin{quote}
$$(\dagger) \hspace{1em} \forall n  \in \mathbb{N}~ \sum_{i=0}^{i=n} i = \frac{n \times (n+1)}{2}$$
\end{quote}   

Consider the following well-formed codata terms:\footnote{This example assumes that the product of codata with data constants defined above in Section \ref{sec:Corecursive-Definitions} is extended to rational numbers.}\footnote{The definitions of \texttt{sum1} and \texttt{sum2} differ in a significant aspect: While \texttt{sum1} is corecursively defined, \texttt{sum2} is \emph{not} corecursively defined.}
\vspace{-2em}
\begin{quote}
\noindent
\begin{verbatim}
(nat, sum1) as [(0, 0)|(1+nat, 1+nat+sum1)]
(nat, sum2) as [(0, 0)|(1+nat, (1/2)*nat*(1+nat))]
\end{verbatim}
\end{quote}

First, a codata predicate $P$ is defined as follows: For all codata names $s$ and $t$, $P(s, t)$ if 
\begin{enumerate}
   \item $s(0) = t(0)$
   \item $\forall n \in \mathbb{N}~ (s(n) = t(n) \Rightarrow s(n+1) = t(n+1))$
\end{enumerate}

Second, it is proved that $P$ is =-hereditary. Let $s$ and $t$ be codata names such that $P(s, t)$ holds.
By (1) of the definition of $P$, $s(0) = t(0)$. 
If follows now from (2) of the definition of $P$ that $s(1) = t(1)$. 
If also follows from (2) of the definition of $P$ that $\forall n \in \mathbb{N} \ {0}~ (s(n) = t(n) \Rightarrow s(n+1) = t(n+1))$. 
This means that $P(s{\myHat}1, t{\myHat}1)$ what completes the proof that P is =-hereditary. 

Third, it is proved that $P$(\verb#sum1, sum2)# holds. 
By definition of \verb#sum1# and \verb#sum2#, \verb#sum1#$(0) =$ \verb#sum2#$(0)$. 
Let \verb#n# $\in \mathbb{N}$ and assume that \verb#sum1#$(n) =$ \verb#sum2#$(n)$. 
By definition of \verb#sum1#:
\begin{quote}
\verb#sum1(n+1) = sum1^(n) + (n+1)#
\end{quote}
By definition of \verb#sum2#:
\begin{quote}
\verb#sum2#$(n+1) = \frac{(n+1) \times (n+2)}{2} = \frac{n \times (n+1)}{2} + (n+1) =$ 
\verb#sum2#$(n) + (n+1)$
 \end{quote}
Thus, $\forall$ \verb#n# $\in \mathbb{N}~ ($\verb#sum1#$(n) =$ \verb#sum2#$(n) \Rightarrow$ \verb#sum1#$(n+1) =$ \verb#sum2#$(n+1))$. 
This completes the proof that $P$\verb#(sum1, sum2)# holds and, by the generalized coinduction proof principle, that of $(\dagger)$.
\end{example}

Under certain notational assumption,  the proof of $\forall n \in \mathbb{N}~ (n > 0 \Rightarrow$ \verb#fib#$(n) > 0)$ given in Example \ref{ex:fib>0} can be shortened as follows. 
From 
\begin{quote}
\verb#fib = [0, 1 | fib + fib^1]#
\end{quote}
it follows that \begin{quote}
\verb#fib^1 as [1, 1 | fib^1 + fib^2]#
\end{quote}
Thus, if \verb#fib^1 #$> 0$ and \verb#fib^2 #$> 0$, then \verb#fib^1 + f^2 #$> 0$. 
Since \verb#fib^1#$(0) =$ \verb#fib^1#$(1) = 1$, the result holds by the generalized coinduction proof principle 

The notational assumption giving sense to the above proof is that if $p$ denotes a data predicate, then the pointwise-derived from $p$, $P_p$, defined in \ref{def:Pointwise-defined-hereditary}, is also denoted $p$, the context allowing to disambiguate. 
The first sentence above states that the expression defining \verb#fib^1#, \verb#fib^1 + fib^2#, preserves the pointwise-derived of $(> 0)$, $P_{(>0)}$. This is indeed the case because so does the ``down-lifted'' of that expression to the data: If $s$ and $t$ are data names such that $s > 0$ and $t > 0$, then $s + t > 0$. 
The second sentence states that each of the finitely many initial values defining \verb#fib^1# does satisfy the data predicate $(>0)$. 

Under the same notational assumption, the proof 
can also be expressed as follows.  
Assume \verb#fib^1 > 0#. \verb#fib^1#$(0) = 1$ and \verb#fib^2#$(1) = 1$ and since \verb#fib^1 > 0#, \verb#fib^1 + fib^2 > 0#. Since there are no contradictions, the assumption \verb#fib^1 > 0# follows by the generalized coinduction proof principle.  

It is similar shortened bisimulation proofs that have made the coinduction proof principle ''seem a bit magical'' \cite[p. 2]{Kozen-Silva-2016}. 

Example \ref{ex:sum} suggests the following theorem. 

\begin{theorem}[Second Generalized Coinduction Theorem]
The principle of mathematical induction and the generalized coinduction proof principle follow from each other. 
\end{theorem}

\begin{proof}
Since the First Generalized Coinduction Theorem \ref{thm:First-Generalized-Coinduction-Theorem}, from which generalized coinduction proof principle is derived, is proved above by induction, it suffices to show that the principle of mathematical induction follows from the generalized coinduction proof principle. 

Assume that $p$ is a unary data predicate such that $p(0)$ holds and $\forall n \in \mathbb{N}~ (p(n) \Rightarrow p(n+1))$. 
We show that the generalized coinduction proof principle implies that $\forall n \in \mathbb{N}~ p(n)$. 

Let $P$ be a codata predicate defined by: 
\begin{quote}
For all codata names $s$,  $P(s)$ holds if
\begin{enumerate}
   \item $p(s(0))$
   \item $\forall n \in \mathbb{N}~ p(s(n)) \Rightarrow p(s(n+1))$
\end{enumerate}
\end{quote}
Let $s$ be a codata name. Assume $P(s))$ holds. it follows from the definition of $P$ that $p(s(1))$ holds and that $\forall n \in \mathbb{N} \setminus \{0\}~ p(s(n)) \Rightarrow p(s(n+1))$, that is, by definition of $P$, $P(s{\myHat}1)$ holds. Thus, $P$ is $p$-hereditary. By the generalized coinduction proof principle, $\forall n \in \mathbb{N}~ p(n)$, what completes the proof. 
\end{proof}

\section{Perspectives and Conclusion}

An observed stream is a possibly never ending sequence of observations or measurements of natural or artificial systems like temperatures, the reproduction rates of a disease, and energy consumption or traffic volumes. 
Observed streams are compared with behavioral models, typically for detecting situations of interest like critical raises of a stream's data.  

Behavioral models of observed streams can be expressed as constructed, or synthetic, streams specified as codata terms (as defined in Sections \ref{sec:Coinductive-Definitions} and \ref{sec:Corecursive-Definitions}). Functions or predicate on codata then give rise to express properties of such models which can be established using the generalized coinduction proof. A proof similar to that of Example \ref{ex:1>0} for example establishes that \verb#s1# $=$ \verb#s2# if \verb#s1# and \verb#s2# are defined by: 
\vspace{-2em}
\begin{quote}
\noindent
\begin{verbatim}
s1 as [1 | s1]
s2 as [1, 1 | s2]
\end{verbatim}
\end{quote}
For a less trivial example, consider the following codata terms which specify repeating oscillations over respectively 2,  3 and 6 states denoted \verb#0# to \verb#5#: 
\vspace{-2em}
\begin{quote}
\noindent
\begin{verbatim}
o2 as [0, 1 | o2]
o3 as [0, 1, 2 | o3]
o6 as [0, 1, 2, 3, 4, 5 | o6]
\end{verbatim}
\end{quote}

\begin{definition}
The codata \verb#s# and \verb#t# as \emph{finitely equivalent}, denoted \verb#s# $\sim_f$ \verb#t#, if 
for some $n \in \mathbb{N}$: 
\begin{enumerate}
   \item{ there is a bijection
   
            \begin{tabular}{l c l c}
               $b:$ 	& $\{\texttt{s}(0), \ldots, \texttt{s}(n)\}$ 	& $\rightarrow$ 	& $\{\texttt{t}(0), \ldots, \texttt{t}(n)\}$ \\
               		& $\texttt{s}(i)$ 					& $\rightarrow$ 	& $\texttt{t}(i)$ 
            \end{tabular}
            }
   \item for all $m \in \mathbb{N}~ \texttt{s}(n+m) = \texttt{s}(n)$ and $\texttt{t}(n+m) = \texttt{t}(n)$
\end{enumerate}
\end{definition} 

Recall that \verb#(o2,o3)# defines (see Section \ref{sec:Corecursive-Definitions}) the stream: \\
\verb#[(0,0), (1,1), (0,2), (1,0), (0,1), (1,2), (0,0), ...]# 
A proof similar to that of Example \ref{ex:1>0} establishes that 
\begin{quote}
\verb#(o2,o3)# $\sim_f$ \verb#o6#
\end{quote}
that is, \verb#(o2,o3)# specifies a stream endlessly oscillating over 6 states.

Consider now the codata term
\begin{quote}
\begin{verbatim}
n as [0 | s(n)]
\end{verbatim}
\end{quote}
which specifies the stream:
\begin{quote}
\begin{verbatim}
[0, s(0), s(s(0)), s(s(s(0))), ...]
\end{verbatim}
\end{quote}
Recall the following codata term from Section \ref{sec:Corecursive-Definitions} which specifies the stream of natural numbers: 
\begin{quote}
\verb#nat as [0 | 1 + nat]#
\end{quote}
A proof by generalized coinduction that \verb# n# $\sim_f$ \verb#nat # does not essentially differ from the proof given in Example \ref{ex:1>0} and from the above-mentioned proof of \verb# (o2,o3)# $\sim_f$ \verb#o6#.

The aforementioned proofs by generalized coinduction have in common that they refer either to structurally defined codata (in the sense of Definition \ref{def:Structural-codata-expression}) or, for the last one, to a decidable first-order theory, Presburger arithmetic \cite{Presburger-Arithmetics-1929}. As a consequence, such proofs can be automated. More precisely, proofs of the finite equivalence (in the sense of $\sim_f$ defined above) of structurally defined codata terms (in the sense of Definition \ref{def:Structural-codata-expression}) and of codata terms defined in terms of decidable theories can be completely automated and therefore supported by a declarative, functional or logic, programming language.\footnote{The technique needed for such an automation is equational unification or e-unification \cite{Plotkin-Building-in-Equational-Theories-1972,Siekmann-Universal-Unification-1984,Degtyarev-Voronkov-What-You-Always-Wanted-To-Know-About-Rigid-E-Unification-1998,Tiwari-Rigid-E-Unification-Revisited-2000,Escobar-Variant-Narrowing-and-Equational-Unification-2009}.}

A declarative programming language providing automated proofs by generalized coinduction for codata terms defined in terms of decidable theories would be a useful tool for processing observed streams. 

This article has first given an extensive review of coinduction in declarative programming. Second, it has reviewed and slightly extended the codata formalism. Third, it has generalized to any predicate  the coinduction proof principle which was originally specified for the equality predicate only. This generalization has been shown to make the coinduction proof principle more intuitive by making better visible its closeness with structural induction. The article has finally suggested to extend declarative, functional or logic, programming with a limited, decidable form of the generalized of the coinduction proof principle based upon structural codata definitions or upon codata definitions referring to decidable theories. 

\section*{Acknowledgments}

The author thanks Schloss Dagstuhl -- Leibniz Rechenzentrum f\"{u}r Informatik and
Alexander Artikis, 
Thomas Eiter, 
Alessandro Margara, 
and 
Stijn Vansummeren, the organisers of the Dagstuhl Seminar  ``Foundations of Composite Event Recognition'' (February 9–14, 2020, \url{https://www.dagstuhl.de/20071}), 
for their invitation to present an early version of the work reported about in this article at the seminar. 

The author thanks the seminar participants for their useful comments on his seminar presentation. 

\bibliography{bry-coinduction-plain-and-simple.bib}

\begin{thebibliography}{}

\bibitem[\protect\citeauthoryear{Abdallah}{Abdallah}{1983}]{Abdallah-Metric-Interpretations-and-Greatest-Fixpoint-Semantics-of-LP-1983}
{\sc Abdallah, N. M.~A.} 1983.
\newblock Metric interpretations and greatest fixpoint semantics of logic
  programs.
\newblock Research Report CS-83-29, Department of Computer Science, University
  of Waterloo, Ontario, Canada.

\bibitem[\protect\citeauthoryear{Abdallah}{Abdallah}{1984}]{Abdallah-Interpretation-of-Infinite-Computations-in-LP-1984}
{\sc Abdallah, N. M.~A.} 1984.
\newblock On the interpretation of infinite computations in logic programming.
\newblock In {\em Proceedings of the 11th International Colloquium on Automata,
  Languages and Programming (ICALP)}. LNCS, vol. 172. Springer, 358--370.

\bibitem[\protect\citeauthoryear{Ancona}{Ancona}{2013}]{ancona-regular-corecursion-in-prolog-2013}
{\sc Ancona, D.} 2013.
\newblock Regular corecursion in {P}rolog.
\newblock {\em Computer Languages, Systems \& Structures\/}~{\em 39,\/}~4,
  142--162.
\newblock Special issue on the Programming Languages track at the 27th ACM
  Symposium on Applied Computing, 2012.

\bibitem[\protect\citeauthoryear{Ancona, Dagnino, and Zucca}{Ancona
  et~al\mbox{.}}{2017}]{Ancona-extending-co-logic-programming-with-cofacts-2017}
{\sc Ancona, D.}, {\sc Dagnino, F.}, {\sc and} {\sc Zucca, E.} 2017.
\newblock Extending coinductive logic programming with co-facts.
\newblock In {\em Proceedings of the First Workshop on Coalgebra, Horn Clause
  Logic Programming and Types}, {E.~Komendantskaya} {and} {J.~Power}, Eds.
  1--18.

\bibitem[\protect\citeauthoryear{Avenhaus, K\"{u}hler, Schmidt-Samoa, and
  Wirth}{Avenhaus et~al\mbox{.}}{2003}]{Avenhaus-Quod-Libet-2003}
{\sc Avenhaus, J.}, {\sc K\"{u}hler, U.}, {\sc Schmidt-Samoa, T.}, {\sc and}
  {\sc Wirth, C.-P.} 2003.
\newblock How to prove inductive theorems? quodlibet!
\newblock In {\em Automated Deduction -- Proceedings of the International
  Conference on Automated Deduction (CADE)}, {F.~Baader}, Ed. LNCS, vol. 2741.
  Springer, 328--333.

\bibitem[\protect\citeauthoryear{Boyer and Moore}{Boyer and
  Moore}{1988}]{Boyer-Moore-Integrating-decision-procedures-into-heuristic-theorem-provers1988}
{\sc Boyer, R.~S.} {\sc and} {\sc Moore, J.~S.} 1988.
\newblock Integrating decision procedures into heuristic theorem provers: A
  case study of linear arithmetic.
\newblock {\em Machine Intelligence\/}~{\em 11}, 83--124.

\bibitem[\protect\citeauthoryear{Bundy}{Bundy}{2006}]{Bundy-automation-of-proof-by-mathematical-induction-2006}
{\sc Bundy, A.} 2001, 2006.
\newblock {\em Handbook of Automated Reasoning}. Vol.~1.
\newblock MIT Press, Chapter The automation of proof by mathematical induction,
  845--912.

\bibitem[\protect\citeauthoryear{Burstall}{Burstall}{1969}]{Burstall-Structural-Induction-1969}
{\sc Burstall, R.~M.} 1969.
\newblock Proving properties of programs by structural induction.
\newblock {\em The Computer Journal\/}~{\em 12,\/}~1, 41--48.

\bibitem[\protect\citeauthoryear{Cantor}{Cantor}{1891}]{Cantor-Diagonalization-1891}
{\sc Cantor, G.} 1891.
\newblock {\"{U}}ber eine elementare {F}rage der {M}annigfaltigkeitslehre.
\newblock {\em {J}ahresbericht der {D}eutschen
  {M}athematiker-{V}ereinigung\/}~1, 75--18.

\bibitem[\protect\citeauthoryear{Church}{Church}{1938}]{Church-Constructive-Second-Number-Class-1938}
{\sc Church, A.} 1938.
\newblock The constructive second number class.
\newblock {\em Bulletin of the American Mathematical Society\/}~{\em 44,\/}~4,
  224--232.

\bibitem[\protect\citeauthoryear{Church and Kleene}{Church and
  Kleene}{1937}]{Church-Kleene-Ordinal-Numbers-1937}
{\sc Church, A.} {\sc and} {\sc Kleene, S.~C.} 1937.
\newblock Formal definitions in the theory of ordinal numbers.
\newblock {\em Fundamenta Mathematicae\/}~{\em 28}, 11--21.

\bibitem[\protect\citeauthoryear{Clark}{Clark}{1978}]{Clark-Negation-as-Failure-1978}
{\sc Clark, K.~L.} 1978.
\newblock {\em Logic and Data Bases}.
\newblock Plenum, New York, USA, Chapter Negation as Failure, 293--324.

\bibitem[\protect\citeauthoryear{Colmerauer}{Colmerauer}{1982}]{colmerauer-prolog-and-infinite-trees-1982}
{\sc Colmerauer, A.} 1982.
\newblock {\em Logic Programming}.
\newblock Academic Press, Cambridge, MA, USA, Chapter Prolog and Infinite
  Trees, 231--251.

\bibitem[\protect\citeauthoryear{Colmerauer}{Colmerauer}{1984}]{colmerauer-equations-inequations-on-infinite-trees-1984}
{\sc Colmerauer, A.} 1984.
\newblock Equations and inequations on finite and infinite trees.
\newblock In {\em Proceedings of the International Conference on Fifth
  Generation Computer Systems (FGCS)}, {I.~for New Generation Computer
  Technology~(ICOT)}, Ed. Ohmsha Ltd. and North-Holland, Tokyo, Japan, and
  Amsterdam, The Netherland, 85--99.

\bibitem[\protect\citeauthoryear{Colmerauer}{Colmerauer}{1988}]{colmerauer-private-communication-1988}
{\sc Colmerauer, A.} 1988.
\newblock Personal communication.

\bibitem[\protect\citeauthoryear{Comon}{Comon}{2006}]{Comon-Inductionless-Induction-2001}
{\sc Comon, H.} 2001, 2006.
\newblock {\em Handbook of Automated Reasoning}. Vol.~1.
\newblock MIT Press, Chapter Inductionless Induction, 913--962.

\bibitem[\protect\citeauthoryear{Courcelle}{Courcelle}{1983}]{courcelle-fundamental-properties-of-infinite-trees-1983}
{\sc Courcelle, B.} 1983.
\newblock Fundamental properties of infinite trees.
\newblock {\em Theoretical Computer Science\/}~{\em 25}, 95--169.

\bibitem[\protect\citeauthoryear{Degtyarev and Voronkov}{Degtyarev and
  Voronkov}{1984}]{Degtyarev-Voronkov-What-You-Always-Wanted-To-Know-About-Rigid-E-Unification-1998}
{\sc Degtyarev, A.} {\sc and} {\sc Voronkov, A.} 1984.
\newblock What you always wanted to know about rigid e-unification.
\newblock {\em Journal of Automated Reasoning\/}~{\em 20}, 47--80.

\bibitem[\protect\citeauthoryear{Escobar, Meseguer, and Sasse}{Escobar
  et~al\mbox{.}}{2009}]{Escobar-Variant-Narrowing-and-Equational-Unification-2009}
{\sc Escobar, S.}, {\sc Meseguer, J.}, {\sc and} {\sc Sasse, R.} 2009.
\newblock Variant narrowing and equational unification.
\newblock {\em Electronic Notes in Theoretical Computer Science\/}~{\em
  238,\/}~3, 103--119.

\bibitem[\protect\citeauthoryear{Fitting}{Fitting}{2002}]{Fitting-Fixpoint-Semantics-for-Logic-Programming-2002}
{\sc Fitting, M.} 2002.
\newblock Fixpoint semantics for logic programming -- a survey.
\newblock {\em Theoretical Computer Science\/}~{\em 278,\/}~1-2, 25--51.

\bibitem[\protect\citeauthoryear{Fokkink}{Fokkink}{2007}]{Fokkink-Introduction-to-process-algebra-1999}
{\sc Fokkink, W.} 1999, 2007.
\newblock {\em Introduction to Process Algebra}.
\newblock Springer.

\bibitem[\protect\citeauthoryear{Friedman and Wise}{Friedman and
  Wise}{1974}]{Friedman-Wise-Tail-Recursion-1974}
{\sc Friedman, D.~P.} {\sc and} {\sc Wise, D.~S.} 1974.
\newblock Unwinding structured recursions into iterations.
\newblock Technical Report~19, Conmputer Science Department Indiana University,
  Bloomington, Indiana, USA.

\bibitem[\protect\citeauthoryear{Friedman and Wise}{Friedman and
  Wise}{1976}]{Friedman-Lazy-evaluation-1976}
{\sc Friedman, D.~P.} {\sc and} {\sc Wise, D.~S.} 1976.
\newblock Cons should not evaluate its arguments.
\newblock In {\em Third International Colloquium on Automata Languages and
  Programming (ICALP)}, {S.~Michaelson} {and} {R.~Milner}, Eds. Edinburgh
  University Press.

\bibitem[\protect\citeauthoryear{Geuvers}{Geuvers}{2009}]{Geuvers-Proof-Assistants-2009}
{\sc Geuvers, H.} 2009.
\newblock Proof assistants: History, ideas and future.
\newblock {\em S\={a}dhan\={a}\/}~{\em 34}, 3--25.

\bibitem[\protect\citeauthoryear{Giovannetti, Levi, Moiso, and
  Palamidessi}{Giovannetti et~al\mbox{.}}{1991}]{Giovanetti-Kernel-LEAF-1991}
{\sc Giovannetti, E.}, {\sc Levi, G.}, {\sc Moiso, C.}, {\sc and} {\sc
  Palamidessi, C.} 1991.
\newblock Kernel-leaf: A logic plus functional language.
\newblock {\em Journal of Computer and System Sciences (JCSS)\/}~{\em 42,\/}~2,
  1349--185.

\bibitem[\protect\citeauthoryear{G\"{o}del}{G\"{o}del}{1931}]{Goedel-Unvollstaendigkeitssaetze-1931}
{\sc G\"{o}del, K.} 1931.
\newblock {\"{U}}ber formal unentscheidbare {S}\"{a}tze der ``{P}rincipia
  {M}athematica'' und verwandter {S}ysteme {I}.
\newblock {\em Monatsheft f{\"u}r Mathematik und Physik\/}~{\em 38}, 173--198.
\newblock In German, reprinted and English translation: \cite[pages
  144-194]{Goedel-Collected-Works-1986}.

\bibitem[\protect\citeauthoryear{G\"{o}del}{G\"{o}del}{1986}]{Goedel-Collected-Works-1986}
{\sc G\"{o}del, K.} 1986.
\newblock {\em {K}urt {G}\"{o}del -- {C}ollected {W}orks}. Vol. I --
  Publications 1929-1936.
\newblock Oxford University Press, Oxford, UK, and New York, USA.
\newblock In German with English translations.

\bibitem[\protect\citeauthoryear{Golson}{Golson}{1988}]{Golson-Declarative-Semantics-for-Infinite-Objects-in-LP-1988}
{\sc Golson, W.~G.} 1988.
\newblock Toward a declarative semantics for infinite objects in logic
  programming.
\newblock {\em Journal of Logic Programming\/}~{\em 5,\/}~2, 151--164.

\bibitem[\protect\citeauthoryear{Gordon}{Gordon}{1994}]{Gordon-Tutorial-Co-induction-and-FunctionalProgramming-1994}
{\sc Gordon, A.} 1994.
\newblock A tutorial on co-induction and functional programming.
\newblock In {\em Functional Programming -- Proceedings of the 1994 Glasgow
  Workshop on Computing}, {K.~Hammond}, {D.~N. Turner}, {and} {P.~M. Sansom},
  Eds. Springer, 78--95.

\bibitem[\protect\citeauthoryear{Gupta, Saeedloei, Vries, Min, Marple, and
  Klu\'{z}niak}{Gupta
  et~al\mbox{.}}{2011}]{Gupta-Infinite-Computation-Co-induction-and-Computational-Logic-2011}
{\sc Gupta, G.}, {\sc Saeedloei, N.}, {\sc Vries, B.~D.}, {\sc Min, R.}, {\sc
  Marple, K.}, {\sc and} {\sc Klu\'{z}niak, F.} 2011.
\newblock Infinite computation, co-induction and computational logic.
\newblock In {\em Proceedings of the 4th International Conference on Algebra
  and Coalgebra in Computer Science (CALCO)}, {A.~Corradini}, {B.~Klin}, {and}
  {C.~C\^{i}rstea}, Eds. LNCS, vol. 6859. Springer, 40--54.

\bibitem[\protect\citeauthoryear{Hansen, Kupke, and Rutten}{Hansen
  et~al\mbox{.}}{2017}]{Hansen-Stream-Differential-Equations-2016}
{\sc Hansen, H.~H.}, {\sc Kupke, C.}, {\sc and} {\sc Rutten, J. J. M.~M.} 2017.
\newblock Stream differential equations: Specification formats and solution
  methods.
\newblock {\em Logical Methods in Computer Science\/}~{\em 13,\/}~1:3, 1--51.

\bibitem[\protect\citeauthoryear{Hanus}{Hanus}{1994}]{Hanus-The-Integration-of-Functions-into-LP}
{\sc Hanus, M.} 1994.
\newblock The integration of functions into logic programming: From theory to
  practice.
\newblock {\em Journal of Logic Programming\/}~{\em 129-20}, 583--628.

\bibitem[\protect\citeauthoryear{Hein}{Hein}{1992}]{Hein-Completions-of-perpetual-logic-programs-1992}
{\sc Hein, J.} 1992.
\newblock Completions of perpetual logic programs.
\newblock {\em Theoretical Computer Science\/}~{\em 99,\/}~1, 65--78.

\bibitem[\protect\citeauthoryear{Henderson and Morris}{Henderson and
  Morris}{1976}]{Henderson-Lazy-evaluation-1976}
{\sc Henderson, P.} {\sc and} {\sc Morris, J.~H.} 1976.
\newblock A lazy evaluator.
\newblock In {\em Conference Record of the Third ACM SIGACT-SIGPLAN Symposium
  on Principles of Programming Languages (POPL)}. Association for Computing
  Machinery, 95--103.

\bibitem[\protect\citeauthoryear{Hoare}{Hoare}{1978}]{Hoare-CSP-1978}
{\sc Hoare, C. A.~R.} 1978.
\newblock Communicating sequential processes.
\newblock {\em Communications of the ACM\/}~{\em 21,\/}~1, 666--677.

\bibitem[\protect\citeauthoryear{Hofstadter}{Hofstadter}{1979}]{Hofstadter-Goedel-Escher-Bach-1979}
{\sc Hofstadter, D.} 1979.
\newblock {\em G\"{o}del, Escher, Bach: An Eternal Golden Braid}.
\newblock Basic Books.

\bibitem[\protect\citeauthoryear{Jacobs and Rutten}{Jacobs and
  Rutten}{1997}]{Jacobs-Rutten-Tutorial-on-CoAlgebras-and-CoInduction-1997}
{\sc Jacobs, B.} {\sc and} {\sc Rutten, J.} 1997.
\newblock A tutorial on (co)algebras and (co)induction.
\newblock {\em Bulletin of the European Association for Theoretical Computer
  Science (EATCS)\/}~{\em 62}, 222--259.

\bibitem[\protect\citeauthoryear{Jaffar}{Jaffar}{1984}]{Jaffar-Efficient-Unification-Over-Infinite-Terms-1984}
{\sc Jaffar, J.} 1984.
\newblock Efficient unification over infinite terms.
\newblock {\em New Generation Computing\/}~{\em 2,\/}~3, 207--219.

\bibitem[\protect\citeauthoryear{Jaffar, Santosa, and Voicu}{Jaffar
  et~al\mbox{.}}{2008}]{Jaffar-Santosa-Voicu-Coinduction-Rule-2008}
{\sc Jaffar, J.}, {\sc Santosa, A.~E.}, {\sc and} {\sc Voicu, R.} 2008.
\newblock A coinduction rule for entailment of recursively defined properties.
\newblock In {\em Proceedings of the 14th International Conference on
  Principles ands Practice of Constraint Programming (CP)}, {P.~J. Stuckey},
  Ed. LNCS, vol. 5202. Springer, 493--508.

\bibitem[\protect\citeauthoryear{Jaffar and Stuckey}{Jaffar and
  Stuckey}{1986}]{Jaffar-Stuckey-Semantics-of-Infinite-Tree-Logic-Programming-1986}
{\sc Jaffar, J.} {\sc and} {\sc Stuckey, P.~J.} 1986.
\newblock Semantics of infinite tree logic programming.
\newblock {\em Theoretical Computer Science\/}~{\em 46,\/}~3, 141--158.

\bibitem[\protect\citeauthoryear{Jaume}{Jaume}{2002}]{Jaume-On-Greatest-Fixpoint-Semantics-of-LogicProgramming-2002}
{\sc Jaume, M.} 2002.
\newblock On greatest fixpoint semantics of logic programming.
\newblock {\em Journal of Logic and Computation\/}~{\em 12,\/}~2, 321--342.

\bibitem[\protect\citeauthoryear{Jeannin, Kozen, and Silva}{Jeannin
  et~al\mbox{.}}{2017}]{CoCaml-2017}
{\sc Jeannin, J.-B.}, {\sc Kozen, D.}, {\sc and} {\sc Silva, A.} 2017.
\newblock {CoCaml}: Functional programming with regular coinductive types.
\newblock {\em Fundamenta Informaticae\/}~{\em 150,\/}~3-4, 347--377.

\bibitem[\protect\citeauthoryear{Kleene}{Kleene}{1938}]{Kleene-Notations-for-Ordinal-Numbers-1938}
{\sc Kleene, S.~C.} 1938.
\newblock On notation for ordinal numbers.
\newblock {\em The Journal of Symbolic Logic\/}~{\em 3,\/}~4, 150--155.

\bibitem[\protect\citeauthoryear{Knaster}{Knaster}{1928}]{Knaster-1928}
{\sc Knaster, B.} 1928.
\newblock Un th\'{e}or\`{e}me sur les fonctions d'ensembles.
\newblock {\em Annales de la Soci\'{e}t\'{e} Polonaise de
  Math\'{e}matique\/}~{\em 6}, 133--134.

\bibitem[\protect\citeauthoryear{Knuth and Bendix}{Knuth and
  Bendix}{1967}]{Knuth-Bendix-Algorithm-1967}
{\sc Knuth, D.~E.} {\sc and} {\sc Bendix, P.~B.} 1967.
\newblock Simple word problems in universal algebra.
\newblock In {\em Proceedings of a Conference Held at Oxford Under the Auspices
  of the Science Research Council Atlas Computer Laboratory}, {J.~Leech}, Ed.
  Pergamon, 263--297.

\bibitem[\protect\citeauthoryear{Komendantskaya and Li}{Komendantskaya and
  Li}{2017}]{Komendantskaya-Productive-Corecursion-in-Logic-Programming-2017}
{\sc Komendantskaya, E.} {\sc and} {\sc Li, Y.} 2017.
\newblock Productive corecursion in logic programming.
\newblock {\em Theory and Practice of Logic Programming\/}~{\em 17,\/}~5-6,
  906--923.

\bibitem[\protect\citeauthoryear{Kozen and Silva}{Kozen and
  Silva}{2016}]{Kozen-Silva-2016}
{\sc Kozen, D.} {\sc and} {\sc Silva, A.} 2016.
\newblock Practical coinduction.
\newblock {\em Mathematical Structures in Computer Science\/}, 1--21.

\bibitem[\protect\citeauthoryear{Kupke, Niqui, and Rutten}{Kupke
  et~al\mbox{.}}{2011}]{Kupke-Stream-Differential-Equations-2011}
{\sc Kupke, C.}, {\sc Niqui, M.}, {\sc and} {\sc Rutten, J.} 2011.
\newblock Stream differential equations: concrete formats for coinductive
  definitions.

\bibitem[\protect\citeauthoryear{Levi and Palamidessi}{Levi and
  Palamidessi}{1988}]{Levi-Palamidessi-Contributions-To-The-Semantics-Of-Logic-Perpetual-Processes-1988}
{\sc Levi, G.} {\sc and} {\sc Palamidessi, C.} 1988.
\newblock Contributions to the semantics of logic perpetual processes.
\newblock {\em Acta Informatica\/}~{\em 25}, 691--711.

\bibitem[\protect\citeauthoryear{Li}{Li}{2018}]{Li-Models-2018}
{\sc Li, Y.} 2018.
\newblock Models of coinductive first-order {H}orn clauses.
\newblock In {\em Proceedings of the 25th Automated Reasoning Workshop (ARW) --
  Bridging the Gap between Theory and Practice}, {M.~Jamnik},
  {A.~Koutsoukou-Argyraki}, {E.~Ayers}, {and} {C.~Mangla}, Eds. University of
  Cambridge, United Kingdom, 8--9.

\bibitem[\protect\citeauthoryear{Lloyd}{Lloyd}{1987}]{Lloyd-Foundations-of-Logic-Programming-1987}
{\sc Lloyd, J.~W.} 1987.
\newblock {\em Foundation of Logic Prograsmming\/}, 2nd ed.
\newblock Springer.

\bibitem[\protect\citeauthoryear{Maher}{Maher}{1988}]{Maher-complete-axiomatization-rational-trees-1988}
{\sc Maher, M.~J.} 1988.
\newblock Complete axiomatizations of the algebras of finite, rational and
  infinite trees.
\newblock In {\em Proceedings of the Third Annual Symposium on Logic in
  Computer Science (LICS)}. IEEE, 348--357.

\bibitem[\protect\citeauthoryear{Mantadelis, Rocha, and Moura}{Mantadelis
  et~al\mbox{.}}{2014}]{Theofrastos-Tabling-Rational-Terms-and-Coinduction-Finally-Together!-2014}
{\sc Mantadelis, T.}, {\sc Rocha, R.}, {\sc and} {\sc Moura, P.} 2014.
\newblock Tabling, rational terms, and coinduction finally together!
\newblock {\em Theory and Practice of Logic Programming (TPLP)\/}~{\em
  14,\/}~Special issue 4-5, 429--443.

\bibitem[\protect\citeauthoryear{Martelli and Rossi}{Martelli and
  Rossi}{1984}]{Martelli-Rossi-Unification-with-Infinite-Trees-1984}
{\sc Martelli, A.} {\sc and} {\sc Rossi, G.} 1984.
\newblock Efficient unification with infinite trees in logic programming.
\newblock In {\em Proceedings of the International Conference on Fifth
  Generation Computer Systems (FGCS)}, {I.~for New Generation Computer
  Technology~(ICOT)}, Ed. Ohmsha Ltd. and North-Holland, Tokyo, Japan, and
  Amsterdam, The Netherland, 202--209.

\bibitem[\protect\citeauthoryear{Milner}{Milner}{1971}]{Milner-Simulation-1971}
{\sc Milner, R.} 1971.
\newblock An algebraic definition of simulation between programs.
\newblock In {\em Proceedings of the 2nd International Joint Conferences on
  Artificial Intelligence (IJCAI)}. British Computer Society, 481--489.

\bibitem[\protect\citeauthoryear{Milner}{Milner}{1980}]{Milner-CCS-1980}
{\sc Milner, R.} 1980.
\newblock {\em A Calculus of Communicating Systems}. LNCS, vol.~92.
\newblock Springer.

\bibitem[\protect\citeauthoryear{Moura}{Moura}{2013}]{Moura-Portable-Efficient-Implementation-of-Coinductive-LP-2013}
{\sc Moura, P.} 2013.
\newblock A portable and efficient implementation of coinductive logic
  programming.
\newblock In {\em Proceedings of the 15th International Symposium on Practical
  Aspects of Declarative Languages (PADL)}. LNCS, vol. 7752. Springer, 77--92.

\bibitem[\protect\citeauthoryear{Mukai}{Mukai}{1983}]{mukai-unification-algorithm-for-infinite-trees-1983}
{\sc Mukai, K.} 1983.
\newblock A unification algorithm for infinite trees.
\newblock In {\em Proceedings of the 8th International Joint Conference on
  Artificial Intelligence (IJCAI), Karlsruhe, West Germany}, {A.~Bundy}, Ed.
  Vol.~1. William Kaufmann, Burlington, MA, USA, 547--549.

\bibitem[\protect\citeauthoryear{Mycielski and Taylor}{Mycielski and
  Taylor}{1976}]{Mycielski-Taylor1976_Article_ACompactificationOfTheAlgebra}
{\sc Mycielski, J.} {\sc and} {\sc Taylor, W.} 1976.
\newblock A compactitication of the algebra of terms.
\newblock {\em Algebra Universalis\/}~{\em 6}, 159--163.

\bibitem[\protect\citeauthoryear{Palamidessi, Levi, and Falaschi}{Palamidessi
  et~al\mbox{.}}{1985}]{Palamidessi-Formal-Semantics-1985}
{\sc Palamidessi, C.}, {\sc Levi, G.}, {\sc and} {\sc Falaschi, M.} 1985.
\newblock The formal semantics of processes and streams in logic programming.
\newblock {\em Colloquia Mathematica Societatis Janos Bolyai\/}~{\em 42},
  363--377.

\bibitem[\protect\citeauthoryear{Park}{Park}{1981}]{Park-Concurrency-on-automata-and-infinite-sequences-1981}
{\sc Park, D.} 1981.
\newblock Concurrency on automata and infinite sequences.
\newblock In {\em Proceedings of the Conference on Theoretical Computer
  Science}, {P.~Deussen}, Ed. LNCS, vol. 104. 167--183.

\bibitem[\protect\citeauthoryear{Paulson}{Paulson}{1997}]{Paulson-Mechanizing-coinduction-and-corecursion-in-higher-order-logic-1997}
{\sc Paulson, L.~C.} 1997.
\newblock Mechanizing coinduction and corecursion in higher-order logic.
\newblock {\em Journal of Logic and Computation\/}~{\em 7,\/}~2, 175--204.

\bibitem[\protect\citeauthoryear{P\'{e}ter}{P\'{e}ter}{1961}]{Peter-Structural-Induction-1961}
{\sc P\'{e}ter, R.} 1961.
\newblock \"{U}ber die {V}erallgemeinerung der {T}heorie der rekursiven
  {F}unktionen f{\"u}r abstrakte {M}engen geeigneter {S}truktur als
  {D}efinitionsbereiche.
\newblock {\em Acta Mathematica Hungarica\/}~{\em 12,\/}~3-4, 271--314.
\newblock In German (``On the generalization of the theory of recursive
  functions with abstract sets of suitable structures as domains''), presented
  at ``Internationales Symposium der Grundlagen der Mathematik: Infinitische
  Methoden'' (``International Symposium on the Foundations of Mathematics:
  Infinistic Methods'') on 3 September 1959 in Warsaw, Poland.

\bibitem[\protect\citeauthoryear{Plotkin}{Plotkin}{1972}]{Plotkin-Building-in-Equational-Theories-1972}
{\sc Plotkin, G.~D.} 1972.
\newblock Building-in equational theories.
\newblock {\em Machine Intelligence\/}, 73--90.

\bibitem[\protect\citeauthoryear{Pous and Sangiorgi}{Pous and
  Sangiorgi}{2011}]{Pous-Sangiorgi-2011}
{\sc Pous, D.} {\sc and} {\sc Sangiorgi, D.} 2011.
\newblock {\em Advanced Topics in Bisimulation and Coinduction}.
\newblock Cambridge Tracts in Theoretical Computer Science. Cambridge
  University Press, Chapter Enhancements of the bisimulation proof method,
  233--289.

\bibitem[\protect\citeauthoryear{Presburger}{Presburger}{1929}]{Presburger-Arithmetics-1929}
{\sc Presburger, M.} 1929.
\newblock {\"{U}}ber die {V}ollst\"{a}ndigkeit eines gewissen {S}ystems der
  {A}rithmetik ganzer {Z}ahlen, in welchem die {A}ddition als einzige
  {O}peration hervortritt.
\newblock In {\em Sprawozdanie z I Kongresu Matematyk\'{o}w Kraj\'{o}w
  (Proceedings of the First Congress of Mathematicians of the Slavic
  Countries)}. Warsaw, Poland, 92--101.
\newblock In German, English translation:
  \cite{Presburger-Arithmetics-1929-English}.

\bibitem[\protect\citeauthoryear{Presburger}{Presburger}{1991}]{Presburger-Arithmetics-1929-English}
{\sc Presburger, M.} 1991.
\newblock On the completeness of a certain system of arithmetic of whole
  numbers in which addition occurs as the only operation.
\newblock {\em History and Philosophy of Logic\/}~{\em 12,\/}~2, 225--232.

\bibitem[\protect\citeauthoryear{Rutten}{Rutten}{2000}]{Rutten-universal-coalgebras-2000}
{\sc Rutten, J. J. . M.~M.} 2000.
\newblock Universal coalgebras: a theory of systems.
\newblock {\em Theoretical Computer Science\/}~{\em 249,\/}~1, 3--80.

\bibitem[\protect\citeauthoryear{Rutten}{Rutten}{2005}]{Rutten-A-Coinductive-Calculus-of-Streams-2005}
{\sc Rutten, J. J. M.~M.} 2005.
\newblock A coinductive calculus of streams.
\newblock {\em Mathematical Structures in Computer Science\/}~{\em 15},
  93--147.

\bibitem[\protect\citeauthoryear{Sangiorgi}{Sangiorgi}{2009}]{Sangiorgi-Origins-of-bisimulation-and-coinduction-2009}
{\sc Sangiorgi, D.} 2009.
\newblock On the origins of bisimulation and coinduction.
\newblock {\em ACM Transactions on Programming Languages and Systems\/}~{\em
  31,\/}~4, 15:1--15.

\bibitem[\protect\citeauthoryear{Sangiorgi}{Sangiorgi}{2012}]{Sangiorgi-Bisimulation-and-Coinduction-2012}
{\sc Sangiorgi, D.} 2012.
\newblock {\em An Introduction to Bisimulation and Coinduction}.
\newblock Cambridge University Press.

\bibitem[\protect\citeauthoryear{Siekman}{Siekman}{}]{Siekmann-Universal-Unification-1984}
{\sc Siekman, J.}
\newblock Universal unification.
\newblock In {\em Proceedings of the 7th Conference on Automated Deduction
  (CADE)}, {R.~E. Shostak}, Ed. LNCS, vol. 170. Springer, 1--42.

\bibitem[\protect\citeauthoryear{Simon, Bansal, Mallya, and Gupta}{Simon
  et~al\mbox{.}}{2006}]{Simon-CoLP-2006}
{\sc Simon, L.}, {\sc Bansal, A.}, {\sc Mallya, A.}, {\sc and} {\sc Gupta, G.}
  2006.
\newblock Coinductive logic programming.
\newblock In {\em Proceedings of the International Conference on Logic
  Programming (ICLP)}, {S.~Etalle} {and} {M.~Truszczy\'{n}ski}, Eds. LNCS, vol.
  4079. Springer, 330--345.

\bibitem[\protect\citeauthoryear{Simon, Bansal, Mallya, and Gupta}{Simon
  et~al\mbox{.}}{2007}]{Simon-CoLP-2007}
{\sc Simon, L.}, {\sc Bansal, A.}, {\sc Mallya, A.}, {\sc and} {\sc Gupta, G.}
  2007.
\newblock Co-logic programming: Extending logic programming with coinduction.
\newblock In {\em Proceedings of the International Colloquium on Automata,
  Languages, and Programming (ICALP)}. LNCS, vol. 4596. Springer, 472--483.

\bibitem[\protect\citeauthoryear{Socher-Ambrosius}{Socher-Ambrosius}{1992}]{Socher-Ambrosius-Ancestor-subsumption-1992}
{\sc Socher-Ambrosius, R.} 1992.
\newblock How to avoid the derivation of redundant clauses in reasoning
  systems.
\newblock {\em Journal of Automated Reasoning\/}~{\em 9}, 77--97.

\bibitem[\protect\citeauthoryear{Steele}{Steele}{1977}]{Steele-Tail-Recursion-1977}
{\sc Steele, G.~L.} 1977.
\newblock Debunking the ``expensive procedure call'' myth or, procedure call
  implementations considered harmful or, lambda: The ultimate goto.
\newblock In {\em Proceedings of the 1977 Annual Conference (ACM)}, {J.~S.
  Ketchel} {and} {H.~Z. Kriloff}, Eds. Association for Computing Machinery, New
  York, USA, 153--162.

\bibitem[\protect\citeauthoryear{Tarski}{Tarski}{1930}]{Tarski-1930}
{\sc Tarski, A.} 1930.
\newblock {\"{U}}ber {\"{a}}quivalenz der mengen in bezug auf eine beliebige
  klasse von abbildungen.
\newblock In {\em Proceedings of the International Congress of Mathematicians
  1928}. Vol.~2. 243--252.

\bibitem[\protect\citeauthoryear{Tarski}{Tarski}{1955}]{Tarski-1955}
{\sc Tarski, A.} 1955.
\newblock A lattice-theoretical theorem and its applications.
\newblock {\em Pacific Journal of Mathematics\/}~{\em 5}, 285--309.

\bibitem[\protect\citeauthoryear{Tiwari, Bachmair, and Ruess}{Tiwari
  et~al\mbox{.}}{2000}]{Tiwari-Rigid-E-Unification-Revisited-2000}
{\sc Tiwari, A.}, {\sc Bachmair, L.}, {\sc and} {\sc Ruess, H.} 2000.
\newblock Rigid e-unification revisited.
\newblock In {\em Proceedings of the 17th International Conference on Automated
  Deduction (CADE)}, {D.~McAllester}, Ed. LNCS, vol. 1831. Springer, 220--234.

\bibitem[\protect\citeauthoryear{van Emden and Abdallah}{van Emden and
  Abdallah}{1985}]{van-Emden-Top-down-semantics-1985}
{\sc van Emden, M.~H.} {\sc and} {\sc Abdallah, N. M.~A.} 1985.
\newblock Top-down semantics of fair computations of logic programs.
\newblock {\em Journal of Logic Programming\/}~{\em 2,\/}~1, 67--75.

\bibitem[\protect\citeauthoryear{van Emden and Kowalsi}{van Emden and
  Kowalsi}{1976}]{van-Emden-Kowalski-Semantics-of-Predicate-Logic-1976}
{\sc van Emden, M.~H.} {\sc and} {\sc Kowalsi, R.~A.} 1976.
\newblock The semantics of predicate logic as a programming language.
\newblock {\em Journal of the ACM\/}~{\em 23,\/}~4, 733--742.

\bibitem[\protect\citeauthoryear{van Emden and Lloyd}{van Emden and
  Lloyd}{1984}]{van-Emden-A-logical-reconstruction-of-PrologII-1983}
{\sc van Emden, M.~H.} {\sc and} {\sc Lloyd, J.~W.} 1984.
\newblock A logical reconstruction of {P}rolog {II}.
\newblock {\em Journal of Logic Programming\/}~{\em 2}, 143--149.

\end{thebibliography}

\end{document}